\documentclass[12pt]{iopart}

\newcommand{\Sh}{Schr\"odinger{ }}

\newcommand{\Fig}[1]{Fig. \ref{#1}}
\newcommand{\braket}[3]{\left\langle #1 \left\vert #2
            \right\vert #3 \right\rangle}
\newcommand{\brakete}[3]{\left\langle #1 \right\vert #2
            \left\vert #3 \right\rangle}
\newcommand{\brak}[2]{\left\langle #1 \left\vert
             #2 \right. \right\rangle}

\newcommand{\fues}[1]{\left(#1\right)}

\newcommand{\yav}[1]{\left[#1\right]}

\newcommand{\llal}[1]{\left\{#1\right.}
\newcommand{\abs}[1]{\left\vert#1\right\vert}

\newcommand{\Eq}[1]{Eq. (\ref{#1})}
\newcommand{\Eqs}[1]{Eqs. (\ref{#1})}

\usepackage{epsfig}  
\usepackage{comment} 
\usepackage{amssymb} 

\def\lambdabar{\protect\@lambdabar}
\def\@lambdabar{%
\relax
\bgroup
\def\@tempa{\hbox{\raise.73\ht0
\hbox to0pt{\kern.25\wd0\vrule width.5\wd0
height.1pt depth.1pt\hss}\box0}}%
\mathchoice{\setbox0\hbox{$\displaystyle\lambda$}\@tempa}%
{\setbox0\hbox{$\textstyle\lambda$}\@tempa}%
{\setbox0\hbox{$\scriptstyle\lambda$}\@tempa}%
{\setbox0\hbox{$\scriptscriptstyle\lambda$}\@tempa}%
\egroup
}

\begin{document}  

\title[Photoconductivity in
AC-driven modulated two dimensional electron gas in a perpendicular
magnetic field ]
      {Photoconductivity in
AC-driven modulated two dimensional electron gas in a perpendicular
magnetic field }    

\author{Manuel Torres}

\address{ Instituto de F\'{\i}sica,
Universidad Nacional Aut\'onoma de M\'exico, Apartado Postal
20-364,  M\'exico Distrito Federal 01000,  M\'exico}
\ead{torres@fisica.unam.mx}

\author{Alejandro Kunold}

\address{Departamento de Ciencias B\'asicas, Universidad Aut\'onoma
Metropolitana Azcapotzalco, Av. San Pablo 180,  M\'exico Distrito Federal  02200, M\'exico}
\ead{akb@correo.azc.uam.mx}

\date{\today}

\begin{abstract}  
 
  In this work we study  the  microwave  photoconductivity   of   a two-dimensional electron system (2DES)
  in the presence   of a magnetic field and a  two-dimensional modulation (2D). The model includes the microwave and Landau contributions in a non-perturbative  exact way, the periodic potential is  treated perturbatively.  The Landau-Floquet states   provide a convenient base with respect to which the lattice  potential becomes  time-dependent,  inducing   transitions between the Landau-Floquet  levels.  Based on this formalism, we provide a  Kubo-like   formula that takes into account the oscillatory Floquet structure of the problem.    The  total  longitudinal conductivity and resistivity  exhibit strong oscillations, determined by
  $\epsilon = \omega /  \omega_c$ with $\omega$ the radiation frequency 
and $\omega_c$ the cyclotron frequency. The oscillations follow a pattern with minima
centered  at $\omega/\omega_c =j + \frac{1}{2} (l-1) +  \delta $,  and maxima centered at  $\omega/\omega_c =j + \frac{1}{2} (l-1) -  \delta $, where $j=1,2,3.......$,   $\delta \sim 1/5$ is a constant  shift and $l$ is the dominant multipole  contribution. 
Negative resistance  states  (NRS) develop as the electron mobility and the  intensity of the microwave power  are increased. These NRS appear in a narrow window region of values of the lattice parameter ($a$),
around $a \sim l_B$, where $l_B$ is the magnetic length. 
It is proposed that these phenomena may be observed  in   artificially   fabricated arrays of periodic scatterers at the interface of  ultraclean  $GaAs/Al_xGa_{1-x} As$  heterostructures.

\end{abstract}


\section{Introduction.}\label{intro}

 The microwave irradiation  of two-dimensional  electron systems $(2DES)$  has remarkable 
consequences on  the transport properties at low magnetic fields.   Recently, two experimental groups
\cite{zudov1,zudov2,mani1,mani2},  reported the unexpected discovery of zero resistance states (ZRS) when  high mobility $GaAs/Al_xGa_{1-x} As$  heterostructures in weak magnetic fields were exposed to millimeter  irradiation.
Unlike the strong magnetic field regime, the Hall resistance is not quantized. 
The magnetoresistance exhibits    
giant  oscillations, periodic in $\epsilon = \omega /  \omega_c$ with $\omega$ the radiation frequency 
and $\omega_c$ the cyclotron frequency;  the series of minima formed at $ \epsilon = j + \delta$, $j=1,2,3....$, $\delta = \frac{1}{2}$  \cite{zudov1,zudov2}, or  $\delta = \frac{1}{4}$ \cite{mani1,mani2}.
These discoveries triggered a large number of experimental  \cite{doro,willett,zudov3,kova,stud}  and  theoretical  \cite{ry1,ry2,durst,andre,shi,lei,vavilov,tor1,tor2,dmi1,dmi2,kennett} studies.  
According to reference \cite{andre}, ZRS  probably originate  from negative resistance states (NRS); 
it was argued   that negative resistance  induces the formation of current domains, yielding  an instability  that  drives  the system  into a  ZRS.  The existence of NRS  was first predicted in the   pioneering work of  Ryzhii \cite{ry1,ry2}. 
Nowadays two distinct mechanisms that produce negative longitudinal conductance are known:   $(i)$ {\it the impurity scattering mechanism}, which is caused by the disorder  assisted absorption and emission of microwaves   
 \cite{ry1,ry2,durst,andre,shi,lei,vavilov,tor1,tor2}, and
$(ii)$ {\it  the distribution function mechanism},  according to which the  microwave absorption modifies  the    electron distribution  function leading to a  negative longitudinal conductance
\cite{doro,dmi1,dmi2,kennett}.  A model  for  the impurity scattering mechanism was proposed previously by the authors  \cite{tor1,tor2}, the model is based on the fact that  the microwave and Landau  dynamics can be exactly taken into account producing well defined Floquet-Landau states.  The disorder effects  are treated perturbatively, inducing transitions between the Floquet-Landau levels.   The model reproduce various of the experimentally observed features, in particular the fact that negative resistance states $(ZRS)$  appear only when the electron   mobility exceeds  a threshold.

 Although the experiments described above do not include 
 the effect of   periodical potential modulations,   exploring its  physical consequences is worthwhile. 
 We can identified at least three reasons for doing so: $(i)$ The study of both weak \cite{weiss-vk} and strong \cite{schuster} periodically modulated 2DES in the presence of magnetic fields has lead  to the discovery of  interesting transport effects, such as  commensurability phenomena and transport anisotropies.  $(ii)$ The theoretical method 
 previously developed in references \cite{tor1,tor2} is well suited to study this kind of system. $(iii)$ The  use of artificially 
 fabricated arrays of periodic scatterers at the interface of  ultraclean  $GaAs/Al_xGa_{1-x} As$  heterostructures   \cite{weiss-vk,schuster,klitbut,vassi} may allow to test  the predictions made by these theoretical studies. 
 
 In this work we make a theoretical study of the  microwave photoconductivity  of a  2DES in the presence 
 of a magnetic field and a  two-dimensional modulation. 
 Two theoretical studies of  modulated 2DES under the combined effects of magnetic and microwave 
 radiation have recently appeared.    Dietel $et. al.$ \cite{dietel}  considered the photoconductivity in the  case of 1D  periodic modulation. The calculation uses first order perturbation theory for both the microwave  field
 as well as for the periodic potential.  Additionally,  the  calculation is restricted by the following conditions:
  the lattice parameter $a$ is small as compared to the cyclotron radius $R_c$, the  temperature $T$  is large with respect to the periodic strength potential $V_0$,  and  $V_0 \ll \hbar \omega_c$.  Due  of the unidirectional structure of the modulation the photocurrents parallel and perpendicular to the modulation  are different. The work of Gumbs  \cite{gumbs} applies for a   strong 2D modulation, but is linear with respect  to the microwave field intensity. His approach made use of  the usual  Kubo formula in which the matrix elements are evaluated   using  the  numerically obtained 
  Hofstadter-type wave functions. 
 In this paper we  address the  case of a 2D modulation under different conditions; in particular we explore the possible appearance of negative resistance states.  
  The Landau and microwave field contributions 
 are exactly taken into account. We analyze the regime in which the following conditions hold:  
 $k \, T \sim  V_0 \le \hbar \omega_c$, and    $\omega \, \tau_{tr}   \sim  \omega_c \, \tau_{tr}  >>  1$. Instead of appealing to the usual Kubo formula,  our approach shows that the  use of the Floquet-Landau   states determines not only the wave functions required to evaluate the  matrix elements; but also leads to a modified Kubo-like formula, in which the oscillations on the density of states induce  the development of negative   resistance states. 
 As a first step we find a unitary transformation that exactly takes into account the dynamics associated with 
 the Landau and radiation contributions. 
  As a second step, the periodic potential is added  perturbatively. With respect to the  Landau-Floquet states, the periodic potential act as  a coherent oscillating field which  induces   transitions between these levels.   Based on this formalism,  we provide  a  Kubo-like  expression for the conductance that incorporates  the   oscillatory Floquet structure of the system.  
 It is found  that both  $\sigma_{xx}$  and $\rho_{xx}$  exhibit strong oscillations determined by 
  $\epsilon = \omega /  \omega_c$. NRS develop for sufficiently high  electron mobility and  strong microwave power. 
   The model is  used to test chirality  effects induced by the magnetic field, calculations are carried out  for various $\mathbf{E}$-field polarization's.  Finally,  we explore the nonlinear regime in which multiple photon exchange  play an essential role, as well as the current-voltage characteristics of the system.

 The paper is organized as follows. In the next section we present
the model and the   method that allow us to obtain the exact solution of the Landau-microwave
system, as well as  the perturbative corrections induced by the periodic  potential. 
 In section \ref{kubo} we develop the formulation of the  linear response theory  valid in arbitrary  magnetic and microwave  fields.    A discussion of relevant numerical calculations is presented in section (\ref{results}). The last section contains a summary of our main results.

\section{The Model.}\label{model}
Let us consider the motion of an  electron in two dimensions subject to a uniform magnetic  field 
 $\mathbf{ B}$  
perpendicular to the plane and a constant electric field $\mathbf{E }_c$, a periodic potential $V$   and driven by  microwave radiation. On the plane the 
 dynamics is governed by the \Sh equation 
\begin{equation}\label{ecs1}
i \hbar \frac{\partial \Psi }{\partial t}= H \Psi  =  \left[  H_{\{B,\omega\}}  +   V ( \mathbf{r} )  \right] \Psi  \, .
\end{equation}
Here   $H_{\{B,\omega\}}$ is written in term of   the covariant derivative 
\begin{equation}\label{ham1}
H_{\{B,\omega\}} = \frac{1}{2m^*} \mathbf{\Pi}^2 \, , \hskip2.0cm   \mathbf{\Pi} =   \mathbf{p}+e 
  \mathbf{A} \, , 
\end{equation}
where $m^*$ is the effective electron mass  over the plane that takes into account the effects of the crystalline atomic structure over the charge carriers. The vector potential  $\mathbf{A}$  includes  all the contributions  of   the   magnetic, electric   and     radiation fields: 
 
 \begin{equation}\label{vecpot}
 \mathbf{A}  = - \frac{1}{2} \mathbf{r} \times \mathbf{B}  +  Re \,\, \left[  \frac{ \mathbf{\epsilon} E_\omega} {\omega} \exp\{ -i \omega t \} \right] \, +  \mathbf{E}^c \, t \,  . 
\end{equation}
  
The  superlattice  potential   $V(\mathbf{r}) $   is  decomposed in a Fourier expansion
\begin{equation}\label{potimp}
 V(\mathbf{r})= \sum_{m \, n}  V_{m \, n} \exp \bigg{\{i}  2 \pi  \left( \frac{m \, x}{a} + \frac{n \,  y}{b}   \right) \bigg { \}} \, .
\end{equation}

     We first consider the exact solution of the microwave driven Landau problem, the periodic potential  effects are  lately added perturbatively. 
     Along this work we shall assume:  $(i)$ a weak modulation    $ \vert  V \vert   \ll \hbar \omega_c $ and   $(ii)$
     the clean limit  $\omega \, \tau_{tr}   \sim  \omega_c \, \tau_{tr}  >>  1$; here $\tau_{tr}$  is the transport  relaxation time that is estimated using its relation to the   electron mobility $\mu = e \tau_{tr} / m^*$.  Based on these  conditions  the use of first order perturbation theory  in $V$ becomes reasonable. Furthermore the second conditions   justifies  the use of a Fermi distribution function evaluated at the Floquet-Landau quasi-energies,  see appendix (\ref{apenB}).
 
 The system  posed by $H_{\{B,\omega\}}$  can be recast as a forced  harmonic oscillator,  a 
 problem that was solved long time ago by Husimi \cite{husi}.  Following the formalism developed in references \cite{Kunold1,Torres1}, we introduce a canonical transformation to new variables 
$Q_\mu,P_\mu$; $\mu=0,1,2$,  according to 
\begin{eqnarray}\label{trancan}
Q_0  &=&  t   \, ,   \hskip5.6cm  P_0 = i \partial_t + e \phi + e  \mathbf{r} \cdot  \mathbf{E}  ,\nonumber \\
\sqrt{e B} Q_1  &=& \Pi_y   \,,  \hskip4.4cm  \sqrt{e B}P_1 =\Pi_x ,\nonumber \\
\sqrt{e B} Q_2  &=&  \Pi_x + e By  \, ,   \hskip3.1cm  \sqrt{e B} P_2= \Pi_y - e B x.
\end{eqnarray}
It is easily verified that the transformation is indeed  canonical, 
the  new variables obey  the commutation rules:  $- \left[Q_0,P_0\right]=  \left[Q_1,P_1\right]= \left[Q_2,P_2\right]=i B$; all other commutators being zero. The inverse transformation gives
$x=  l_B  \left(Q_1-P_2 \right)$, and  $ y=  l_B   \left( Q_2-P_1 \right) $, 
where $l_B=  \sqrt{\frac{\hbar}{eB}}$ is      the magnetic length. 
 The operators $( Q_2 , P_2)$ can  be  identified with the generators of the electric-magnetic translation symmetries \cite{Ashby1,Kunold2}. Final results are independent  of the selected gauge. 
From the operators in \Eq{trancan}   we construct two pairs of
harmonic oscillator-like ladder operators: $(a_1,a_1^{\dag})$, and $(a_2,a_2^{\dag})$ with:
\begin{equation}\label{opasc}
a_1 =   \frac{1}{\sqrt{2}}    \left(P_1-iQ_1\right), \hskip2.0cm    a_2 =  \frac{1}{\sqrt{2} }    \left(P_2 -i Q_2\right), 
\end{equation}
obeying:  $[a_1,a_1^{\dag}]=[a_2,a_2^{\dag}]=1$, and $[a_1,a_2]=[a_1,a_2^{\dag}]=0$.

It  is now  possible to find a unitary transformation that exactly diagonalizes    $H_{\{B,\omega\}}$, it yields
\begin{equation}\label{tran1}
W^{\dagger} H_{\{B,\omega\}}  W = \omega_c \left( \frac{1}{2} + a_1^\dagger \, a_1\right) \equiv H_0 \, , 
\end{equation}
with the cyclotron frequency $\omega_c = eB/m^*$ and the $W(t)$ operator     given by 
\begin{equation}\label{opw}
\hskip-1.0cm W(t)=  \exp\{i \eta_1 Q_1\}  \exp\{i \xi_1 P_1\} \exp\{i \eta_2 Q_2\}  \exp\{i \xi_2 P_2\}   \exp\{i  \int^t  {\mathcal L} dt^\prime \} 
\, , 
\end{equation}
where  the functions $\eta_i(t)$ and $\xi_i(t)$ represent the solutions to the classical equations of motion that follow from the variation of the Lagrangian 
\begin{equation}\label{lagrang}
{\mathcal L}  =  \frac{\omega_c}{2} \left( \eta_1^2 + \zeta_1^2 \right) +    \dot \zeta_1\eta_1 +  \dot \zeta_2 \eta_2 
+  e l_B \,    \left[ E_x  \left( \zeta_1 + \eta_2 \right)   +    E_y   \left(\eta_1 + \zeta_2  \right) \right]     \, .
\end{equation}
 It  is  straightforward  to obtain the solutions to the equation of motion, using  the expression for the electric  field 
 $ \mathbf{E} = -\partial \mathbf{A} / \partial t  $ with $ \mathbf{A}$ given   in  (\ref{vecpot}). Adding   a damping term that takes into account the radiative decay of the quasiparticle, they read 
 \begin{eqnarray}\label{soleqm}
\hskip-2.0cm & & \eta_1 =  e l_B  E_\omega  \, Re \left[\frac{-i \omega \epsilon_x + \omega_c \epsilon_y  }{\omega^2 - \omega_c^2 + i \omega \Gamma_{rad} }
     e^{i \omega t} \right]  ,   \hskip0.8cm     \eta_2 = e l_B E_\omega \,  Re \left[\frac{\epsilon_y e^{i\omega t}}{i \omega} \right] + e l_B E^c_y  \, t  ,                  \nonumber \\
\hskip-2.0cm & & \zeta_1 = e l_B  E_\omega \, Re \left[\frac{ \omega_c \epsilon_x + i \omega \epsilon_y  }{\omega^2 - \omega_c^2 + i \omega \Gamma_{rad}  }
     e^{i \omega t} \right] ,  \hskip0.8cm      \zeta_2 = -e l_B E_\omega \, Re \left[\frac{\epsilon_x e^{i\omega t}}{i \omega} \right] 
     - e l_B E^c_x  \, t   .
\end{eqnarray}
According to the Floquet theorem,  the wave function  can be written as
 $\Psi (t) = \exp \left(- i {\cal E} _\mu  t \right) \phi_\mu(t)$, where $ \phi_\mu(t)$ is periodic in time, 
 $i.e.$ $ \phi_\mu(t + \tau_\omega ) =  \phi_\mu(t)$, with $\tau_\omega = 2 \pi /\omega$.  From   \Eq{opw} it is noticed that  the transformed wave function  $\Psi^W = W \Psi$ contains the phase factor $\exp \left( i \int^t   {\mathcal L}  d t^\prime \right) $. It then follows  that the quasienergies and the  Floquet modes can be deduced if we add and subtract to this  exponential  a term of the form $\frac{t}{\tau_\omega} \int_0^{\tau_\omega} {\mathcal L} dt^\prime$. Hence, the quasienergies can be  readily read off
\begin{equation}\label{enerflo}
\hskip-2.0cm  {\cal E} _\mu ={\cal E} _\mu^{(0)} +   {\cal E}_{rad} \, ;  \hskip0.6cm  
   {\cal E} _\mu^{(0)} = \hbar \omega_c \left( \frac{1}{2} + \mu  \right) ,   \hskip0.6cm  {\cal E}_{rad} = 
   \frac{e^2 E_\omega^2 \left[ 1 + 2 \omega_c {Re(\epsilon_x^* \epsilon_y) }/ \omega\right] } {2 m^* \left[ \left(\omega - \omega_c \right)^2 + \Gamma_{rad}^2 \right]}, 
\end{equation}
here  ${\cal E} _\mu^{(0)}$ are   the usual Landau energies, and the induced  Floquet energy shift is given by the microwave energy 
$ {\cal E}_{rad} $.   
The corresponding time-periodic Floquet modes in the 
  $(P_1,P_2)$ representation are given by 
\begin{equation}\label{wf2}
\Psi_{\mu,k } (P) = \exp\{- i \sin\left(2 \omega t \right) F(\omega) \} \phi_\mu(P_1) \delta(P_2 - k) \, , 
\end{equation}
the  index $k$ labels the degeneracy of the Landau-Floquet states, and  $\phi_{\mu}\fues{P_1}$ is the harmonic oscillator function  in the 
$P_1$ representation
\begin{equation}\label{hofun}
\phi_{\mu}\fues{P_1}=\brak{P_1}{\mu}=
\frac{1}{\sqrt{\pi^{1/2}2^{\mu}\mu !}}
e^{-P_1^2/2}H_{\mu}\fues{P_1} \, , 
\end{equation}
 $H_{\mu}\fues{P_1}$  is the Hermite polynomial and the function $F(\omega)$ is given as 
\begin{equation}
\hskip-2.5cm F(\omega) = \frac{\omega_c}{\omega}\left(\frac{e E_\omega l_B }{\omega^2 - \omega_c^2 } \right)^2 \left[  \omega^2 - \omega_c^2  + 2 \omega^2 \epsilon_x^2
- 2 \omega_c^2  \epsilon_y ^2 + \frac{Re(\epsilon_x^* \epsilon_y) }{\omega \omega_c}  \left(2 \omega^4 - \omega^2  \omega_c^2 
+ \omega_c^4 \right)  \right]. 
\end{equation}

Let us now consider the complete Hamiltonian including the contribution from the periodic potential. 
 When  the transformation induced by  $W(t)$ is applied,   the   Schr\"odinger equation in (\ref{ecs1})  becomes
\begin{equation}\label{ecs2}
P_0 \Psi^{(W)}  =   H_0 \Psi^{(W)} +   V_W (t) \Psi^{(W)} \, , 
\end{equation}
where $\Psi^{(W)}= W(t)  \Psi$ and  $V_W  (t)  = W(t)  V (\mathbf{r}) W^{-1}(t)$.      Notice that the periodic  potential acquires a time dependence  brought by   the  $W(t)$ transformation. The problem is now solved in the interaction representation  using first order time dependent perturbation theory. 
In the interaction representation $\Psi^{(W)}_I = \exp\{i H_0 t\} \Psi^{(W)}$, and the  Schr\"odinger equation becomes 
\begin{equation}\label{ecs3}
i \partial_t  \Psi^{(W)}_I  =  \left \{ V_W (t) \right \}_I  \Psi^{(W)}_I  \, . 
\end{equation}
The equation has the solution  $\Psi^{(W)}_I  (t) = U(t - t_0) \Psi^{(W)}_I  (t_0)$, where $U(t)$  is the evolution operator. To first order in perturbation theory it  is given by the expression 
\begin{equation}\label{opu}
U(t)  = 1 - i \int_{-\infty}^{t}  dt^\prime \left[ W^{\dagger}(t^\prime) V (\mathbf{r}) W(t^\prime)   \right]_I  \, .
\end{equation}
 The interaction is adiabatically  turned off as $t_0 \to  - \infty$,
in which case the asymptotic  states are selected as   the Landau-Floquet  eigenvalues of $H_0$, $i.e.$ 
$\vert  \Psi^{(W)}_I (t_0) \rangle \to  \vert  \mu,k \rangle$.
Utilizing the explicit expression for the $W$ transformation in (\ref{opw}) and after a lengthly calculation the matrix elements of 
the evolution operator can be worked out as   
  \begin{equation}\label{meU}
\hskip-2.5cm  {\brakete{ \mu,k}{U(t)  }{ \nu,k^\prime}} = \delta_{\mu\nu} \delta_{kk^\prime} - \sum_l \sum_{mn} 
   \delta \left( k - k^\prime + l_B q_n^{(y)}  \right)
  \frac{e^{il_B q_m^{(x)} (k+l_Bq_n^{(y)}/2)} e^{i \left( {\cal E} _{\mu\nu} + \omega l \right)t}}{{\cal E} _{\mu\nu} + \omega l   + \omega_E  } \, C^{(l)}_{\mu\nu,mn},
\end{equation}
  where $\omega_E = e l_B^2 (q^{(y)}_n E^c_x - q^{(x)}_m E^c_y)$, 
 and the explicit expression for $ C^{(l)}_{\mu\nu,mn}$ is  given by  
\begin{equation} \label{defc}
 C_{\mu,\nu,mn}^{(l)}   =  \, l_B^2   \,    V_{m \, n} \,   D_{\mu\nu}(\tilde{q}_{mn} ) \,  \left(\frac{\Delta_{m \,n} }{i \vert  \Delta_{m \, n}  \vert} \right)^l \, 
 J_l  \left( \vert  \Delta_{m \, n}  \vert \right) \, .
\end{equation}
In the previous expressions the    discreet pseudomomentum are given as
\begin{eqnarray} \label{psemom}
 q^{(x)}_m &=& 2 \pi m /a \, ,  \hskip4.0cm q^{(y)}_n= 2 \pi n /b \, ,  \nonumber \\
 \tilde{q}_{mn} &=& i l_B (q^{(x)}_m - i q^{(y)}_n )/ \sqrt{2},
\end{eqnarray}
  $J_l $  denote  the   Legendre polynomials  and   $D_{\mu \nu} $ is given in terms of the  generalized  Laguerre polynomials according to 
\begin{equation}\label{laguerres}
\hskip-1.0cm D^{\nu \mu}\fues{ \tilde{q}}=\braket{\nu}{ D\fues{ \tilde{q}}}{\mu}
=e^{-\frac{1}{2}\abs{  \tilde{q}}^2}
\llal{\begin{array}{ll}\fues{- \tilde{q}^{*}}^{\mu-\nu}
\sqrt{\frac{\nu!}{\mu!}}L^{\mu-\nu}_{\nu}
\fues{\abs{ \tilde{q}}^2}, \,\,\,\,& \mu >\nu, \\
 \tilde{q}^{\nu-\mu}\sqrt{\frac{\mu!}{\nu!}}
L^{\nu-\mu}_{\mu}\fues{\abs{ \tilde{q}}^2},
& \mu <\nu,\\
\end{array}}
\end{equation}
and 
\begin{equation} \label{defdelta2}
\hskip-1.0cm  \Delta_{m \, n}  =  \frac{\omega_c l_B^2 e E_\omega }{\omega \left( \omega^2 - \omega_c^2 + i \omega \Gamma_{rad}  \right)}
\left[ \omega \left(q^{(x)}_m e_x + q^{(y)}_n e_y  \right)  + i \omega_c \left(q^{(x)}_m e_y  -  q^{(y)}_n e_x  \right) \right] \, . 
\end{equation}
Summarizing,  the solution to the original \Sh equation in \Eq{ecs1} has been achieved 
by means os three successive  transformations:
\begin{equation}\label{3trans}
\vert  \Psi_{\mu,k} (t) \rangle  = W^\dag  \,  \exp\{-i H_0 t\} \,  U(t - t_0) \,  \vert  \mu ,k \rangle ,
\end{equation}
 the  explicitly  expressions for $H_0$, $W$, and $U$ are given in Eqs. (\ref{tran1}), (\ref{opw}), and (\ref{meU}) respectively. 

 \section{Kubo formula for Floquet states.}\label{kubo}

The usual Kubo formula for the conductivity must be modified in order to include the Floquet dynamics.
In the presence of an additional $DC$ electric field  the complete Hamiltonian is 
 $H_T = H + V_{ext},$
where $H$ is the Hamiltonian in \Eq{ecs1} and $V_{ext} = \frac{1}{m} {\mathbf{\Pi} } \cdot  \mathbf{A}_{ext}$, with  $\mathbf{A}_{ext}= \frac{ \mathbf{E}_0}{\omega} \, sin \left( \Omega t\right) \, exp \left( - \eta \vert t \vert  \right).$ 
The static limit is obtained with $\Omega \to 0$, and $\eta $ represents the rate at which the perturbation is turned on and off. In order to calculate the expectation value of the current density, we need the density matrix  $\rho(t)$  which obeys  the von Neumann equation 
\begin{equation}\label{vn1}
i \hbar \frac{\partial \rho  }{\partial t}  =  \yav{H_T , \rho} =  \yav{H + V_{ext}  , \rho}.
 \end{equation}
We write to first order   $\rho = \rho_0 + \Delta \rho$, where the leading term 
 satisfies the equation  
 \begin{equation}\label{vn0}
  i \hbar \frac{\partial \rho_0  }{\partial t}  =  \yav{H , \rho_0} \, . 
  \end{equation}
In agreement  with  \Eq{3trans},      $\Delta \rho$ is transformed to 
\begin{equation} \label{dmtran}
\hskip-1.5cm  \tilde {\Delta \rho}  (t)   = U^\dag_I(t-t_0)   \exp\{i H_0 t\}  W(t)  \, {\Delta \rho(t)} \,  W^\dag (t)  \exp\{-i H_0 t\}   U_I(t-t_0) \, . 
\end{equation}
In terms of the transformed  density matrix $  \tilde {\Delta \rho}  (t) $,  \Eq{vn1} becomes
 \begin{equation}\label{vn2}
  i \hbar \frac{\partial   \tilde {\Delta \rho}  }{\partial t}  =  \yav{\tilde{V}_{ext} , \tilde{ \rho}_0 },  
   \end{equation}
where $\tilde{V}_{ext}$ and $ \tilde{ \rho}_0$ are the    external potential and quasi-equilibrium density matrix 
transformed  in the same manner as  $\tilde {\Delta \rho}$ in  \Eq{dmtran}. The transformed  
quasi-equilibrium density matrix is assumed to have the form  
$ \tilde{ \rho}_0 = \sum_\mu \vert \mu \rangle f({\cal E}_\mu) \langle \mu \vert ,$ where $f({\cal E}_\mu)$  is the usual Fermi function and ${\cal E}_\mu$ the Landau-Floquet levels. 
The justification for selecting a  Fermi-Dirac distribution in the quasi-energy states is presented in the appendix (\ref{apenB}).   If we consider the region in which the  conditions 
$\tau_\omega \ll \tau_{tr} \ll \tau_{in} $  hold, then  the  elastic and inelastic relaxation processes can be neglected as compared to the microwave  field effects. The solution of the  Boltzmann equation  yields,   a Fermi-Dirac distribution in the quasi-energy states  \cite{tor1},  see appendix (\ref{apenB}).
It is straightforward to verify that  this selection guarantee that the 
quasi-equilibrium condition in  (\ref{vn0}) is verified.  Using the results in Eqs. (\ref{3trans})  and   (\ref{dmtran}), 
the  expectation value  of the density matrix   can now  be  easily obtained from the integration of \Eq{vn2} 
  with the initial  condition $ {\Delta \rho}  (t)  \to 0$ as $t \to -\infty$ giving for $t < 0$ 
\begin{eqnarray}\label{emdm}
\hskip-2.2cm  {\brakete{\Psi_{\mu,k}}{ {  {\Delta \rho}  (t) } }{\Psi_{\nu,k^\prime}}}     &=&
  {\brakete{\mu,k}{ { \tilde {\Delta \rho}  (t) } }{\nu,k^\prime}}        \nonumber \\
&=&  \frac{e \mathbf{E}_0}{2} \cdot  \int_{-\infty}^t \left[    
  \frac{e^{i(\Omega - i \eta) t^\prime}}{\Omega}  f_{\mu \nu}  {\brakete{\Psi_{\mu,k}}{  \mathbf{\Pi} (t^\prime)}{\Psi_{\nu,k^\prime}}}  + \left(\Omega \to - \Omega \right) \right],
\end{eqnarray}
where  the definition  $f_{\mu \nu} =  f({\cal E}_\mu) - f({\cal E}_\nu)$ was used.
The  expectation value for the momentum operator  is explicitly computed with the help  of Eqs   (\ref{opw}),  (\ref{meU}), and   (\ref{3trans}),  retaining  terms linear in the modulation potential and after a lengthly calculations it yields 
  \begin{equation}\label{emopm}  
\hskip-2.5cm   {\brakete{\Psi_{\mu k}}{ \mathbf{\Pi}_i}{\Psi_{\nu k^\prime}}} =    \, \sqrt{e B} \,  \sum_l \sum_{mn} \, 
   \delta \left( k - k^\prime + l_B q_n^{(y)} \right) e^{il_B q_m^{(x)} (k+l_Bq_n^{(y)}/2)} \,  e^{i ( {\cal E} _{\mu \nu} + \omega l - i \eta)  t} \, \, 
  \Delta ^{(l)}_{\mu \nu,mn} (j) .
 \end{equation} 
Here the following definitions were introduced:  $ {\cal E} _{\mu \nu} =  {\cal E} _\mu - {\cal E} _\nu$,  $a_j = b_j= 1 $ if $j=x$,  $a_j = -b_j= -i $  if $j=y$,  and   $ \Delta ^{(l)}_{\mu \nu,mn} (j)$ is given by 
\begin{equation} \label{defdelta}
  \Delta ^{(l)}_{\mu \nu,mn} (j) =
  -  \frac{1}{\sqrt{2}} \left[  \frac{  a_j \tilde{q}_{mn}^* C_{\mu\nu,mn}^{(l)}}{{\cal E} _{\mu\nu} - \omega_c + \omega l - i \eta }
 +   \frac{ b_j \tilde{q}_{mn} C_{\mu\nu,mn}^{(l)}}{{\cal E} _{\mu\nu} + \omega_c + \omega l - i \eta }\right],
\end{equation}
the expression for $C^{(l)}_{\mu\nu,mn}$ is given in (\ref{defc}). It should be pointed out that in principle there is  a zero order  contribution (independent of  $V$) to the expectation value of the momentum operator in (\ref{emopm}); this would contribute to the direct cyclotron resonance heating that has a single peak around $\omega \sim \omega_c$. However this contribution has proved to be negligible \cite{tor1}, so it will be altogether omitted. 
Utilizing these results the time  integral in \Eq{emdm} is readily carried out. 
 The current density to first order in the external electric field  can now be calculated from 
  $\langle  \mathbf{J} (t,  \mathbf{r}) \rangle  = Tr \left[   \tilde {\Delta \rho}  (t)   \tilde \mathbf{J} (t)  \right] $, the resulting expression represents the local      density current. Here we are concerned with the macroscopic  conductivity tensor that relates the spatially and time averaged  current density  
  $  \mathbf{j} = \left( \tau_\omega {\cal A}  \right)^{-1} \int_0^{\tau_\omega} dt \int d^2x \langle  \mathbf{J} (t,  \mathbf{r}  ) \rangle$ to the averaged electric field; here  ${\cal A}$ is the area of the system (it is understood that ${\cal A} \to \infty)$.
  Assuming that the external electric field points along the $x$-axis the macroscopic conductivity  can be worked out. The total conductivity is given by a sum 
  $\sigma_{xi} = \sigma_{xi}^{D}  + \sigma_{xi}^{(MM)}$;   the  dark  conductivity is  calculated  when both  the  modulation and   microwave radiation are switched-off as 
\begin{equation}\label{condd}
\hskip-1.5cm \mathbf{\sigma}^{D}_{xi}  =   \frac{e^2 \omega_c^2}{4 \hbar \, i } \sum_{\mu \nu} \left\{    \frac{ f_{\mu\nu}}{\Omega}
  \left[ \frac {a_i \mu \delta_{\mu , \nu +1} } {{\cal E}_{\mu\nu} +\Omega - i \eta } 
  + \frac {b_i \nu \delta_{\mu , \nu -1} } {{\cal E}_{\mu\nu} +\Omega - i \eta }  \right]        + \left(\Omega \to - \Omega \right)   \right\},
 \end{equation}
whereas  the microwave-modulation (MM)  induced conductivity is worked out as 
\begin{equation}\label{condw}
\hskip-1.5cm \mathbf{\sigma}^{(MM)}_{xi}    =   \frac{e^2 \omega_c^2}{4 \hbar \, i} \sum_{\mu \nu} \left\{    \frac{ f_{\mu\nu}}{\Omega}
 \sum_{mn}  \sum_l \frac{\Delta_{\mu\nu,mn}^{(l)} (i) \Delta_{\nu\mu,mn}^{(-l) }(x) }{ {\cal E}_{\mu\nu} + \omega l + \Omega - i \eta}
   + \left(\Omega \to - \Omega \right) \right\}.
 \end{equation}
   Selecting $i =x$ or $i=y$  the 
 longitudinal and Hall conductivities  can be selected. The denominators on the R.H.S. of the previous equations 
 can be related to the advanced and  retarded  Green's functions   $G^{\pm}_\mu ({\cal E} ) = 1/\left({\cal E}  -{\cal E} _\mu  \pm i \eta  \right) $. To make further progress the real and absorptive parts of the 
 Green's functions are separated  taking the limit  $\eta \to 0$  and using 
 $lim_{\eta \to 0} \, 1/({\cal E}  -  i \eta) = P 1/{\cal E}   + i \pi \delta({\cal E} )$, where $P$ indicates the principal-value integral. As
 usual  the real and imaginary  parts contribute to the Hall and  longitudinal conductivities respectively.
  However, the  previous expression  would present a singular  behavior that is  an artifact of the $\eta \to 0$ limit. This problem is solved by including the  disorder broadening effects.
    A formal procedure to obtain the Green function     requires a self-consistent 
    calculation using the Dyson equation for the self-energy with the magnetic and microwave fields, impurity, phonon, and other scattering effects included. 
     A  detailed calculation of  $Im \, G_\mu ({\cal E} ) $  incorporating  all these elements is beyond the scope of the present  work. Instead we choose a gaussian-type expression for the  the density of states.  This expression can be justified   within a 
   self-consistent Born  calculation  that incorporates the   magnetic field and  disorder  effects  \cite{ando1,gerh,ando2},   hence  the  density of states for the $\mu$-Landau level   is represented  as 
    \begin{equation} \label{denst}
  Im \, G_\mu({\cal E} ) = \sqrt{\frac{\pi}{2 \Gamma_\mu^2}}  \exp{\left[ - ({\cal E}  - {\cal E} _\mu)^2 /(2 \Gamma_\mu^2) \right] } , 
 \end{equation}   
 with a broadening width given by 
 \begin{equation} \label{denst2}
\hskip1.5cm    \Gamma^2_\mu = \frac{ 2\beta_\mu  \hbar^2 \omega_c  }{ (\pi \tau_{tr})}   , 
 \end{equation}   
the parameter  $\beta_\mu $   takes into account the difference  of the transport scattering time $\tau_{tr}$  determining the mobility $\mu$,  from the single-particle  lifetime $\tau_s$. In the case of short-range scatterers  $\tau_{tr} = \tau_s$ and  $\beta_\mu =1 $. An expression for $\beta_\mu$,  suitable for numerical evaluation,   that applies for a long-range screened potential  is given in
reference \cite{tor1};  $\beta_\mu$  decreases for higher Landau levels; $e.g.$ $\beta_0 \approx 50$, $\beta_{50}  \approx 10.5$.    
 
The static  limit with respect to the external field is obtained taking 
   $\Omega \to 0$   in \Eqs{condd} and (\ref{condw}).  In what follows  results are presented  for the   microwave-modulation (MM)  induced  longitudinal conductivity, the dark conductivities  as well as the MM Hall  conductivity  are quoted  in the appendix.   
 Hence the MM induced longitudinal conductance is worked out as  
 \begin{equation}\label{condLw}
\hskip-2.5cm  \mathbf{\sigma}_{xx} ^{(MM)}   =   \frac{ e^2  l_B^2 }{\pi \hbar}  
\int d {\cal E}  \sum_{\mu \nu}  \sum_l  \,   \sum_{m \, n}  \, Im G_\mu \left({\cal E}   \right)B^{(l)}   \left({\cal E}  ,{\cal E} _\nu \right)       \,
  \bigg{ \vert} q_n^{(y)} \,  J_l\left( \vert \Delta_{m \, n}  \vert \right)  V_{m \, n}   D_{\mu\nu} (\tilde{q}_{mn}) \bigg{\vert}^2  ,
 \end{equation}
 where   the following  function has been defined   
 \begin{equation} \label{derspec} 
\hskip-2.0cm  B^{(l)} ( {\cal E} ,  {\cal E} _\nu) =  -  \frac{d}{d{\cal E} _0}  \bigg{ \{ } \left[ f( {\cal E} + l \omega + \omega_E + {\cal E} _0)-  f ( {\cal E} )\right]  
 Im \, G_\nu ( {\cal E}  + l \omega  + \omega_E  + {\cal E} _0) \bigg{ \}}
  \bigg{\vert}_{{\cal E} _0= 0}. 
 \end{equation}
  Notice that   $\sigma_{xx}^{(MM)}$  contains a contribution  $\sigma_{xx}^{(M)}$   that depends only on   the modulation potential, it  can be extracted from  \Eq{condLw} if the microwave field is switched-off. 
 
  As usual the resistivities are obtained from the expression 
   $ \rho_{xx} =\sigma_{xx}/\left(  \sigma_{xx}^2 + \sigma_{xy}^2 \right)$ and  $ \rho_{xy} =\sigma_{xy}/ \left( \sigma_{xx}^2 + \sigma_{xy}^2 \right)$. The relation $\sigma_{xy} \gg \sigma_{xx}$ holds in general,  hence it follows that $\rho_{xx} \propto \sigma_{xx}$,  and the longitudinal resistivity  follows the same oscillation  pattern as  that of  $\sigma_{xx}$.        
 
 \section{Results.}\label{results}
 The expression in \Eq{condLw}  can be numerically evaluated after the  Fourier components $ V_{m \, n}$ of the periodic  potential    are  specified.   We shall consider a square lattice potential of the form 
   \begin{equation}  \label{pote1} 
  V(\mathbf{r}) = V_0 \left[ \cos\left( \frac{2 \pi x }{a} \right)  +  \cos\left( \frac{2 \pi y }{a} \right) \right]  \, .
 \end{equation}

In our calculations it is assumed  that a  superlattice  is cleaved   at the interface of an  ultraclean  $GaAs/Al_xGa_{1-x} As$ heterostructure   with high  electron  mobility,  $ \mu \sim 0.5-2.5 \times 10^7 cm^2/V  s$; the periodic  potential  has the form given in  (\ref{pote1}) with parameters    $a \sim 20 -200 \, nm$  and  $V_0=0.05 \, meV $. The other parameters  of the sample  are estimated as  effective  electron mass $m^* = 0.067 \, m_e$,       fermi energy $\epsilon_F = 10 \, meV$, electron density $n = 3 \times 10^{11} cm^{-2}$,  and temperature $T =1 \,  K$.  For  the applied external fields we consider:   magnetic fields in the range $0.05 - 0.4 \, \, Tesla$,  and  microwave radiation with frequencies $f  \sim 10-200 \, Ghz$; 
and  field intensity   $ \vert \vec E_\omega \vert  \sim 1-100 \, V/cm$,  that corresponds to a microwave power  characterized by the dimensionless quantity  $\alpha \sim {c \epsilon_0 \vert  E_\omega\vert ^2}/{(m^* \omega^3)}$ that varies in the range $\alpha \sim 0.01 -2$.  The relaxation time $\tau_{tr}$  in \Eq{denst}  is related with  the zero field  electron mobility  through 
   $\mu = e \tau_{tr} /m^*$,  and $\beta_\mu  \approx 10.5$, a value that is justified for large filling factors   $\mu \approx 50$ \cite{tor1}.  A detailed account of the electron dynamics requires to distinguish  between  various 
   time life's; following  reference \cite{mikha},   $\Gamma_{rad}$ in \Eq{soleqm}  is related to 
   the   radiative decay  width   that is interpreted as coherent dipole re-radiation of  electromagnetic waves by the oscillating 2D
   electrons excited by microwaves, it is given by 
    $\Gamma_{rad} = 2 \pi^2 \hbar n e^2 /\left( 3 \epsilon_0 c \, m^* \right)$. 
    In all the  examples, except in    \Fig{figure8}, we consider the linear regime, 
     the dc-electric field is included only through the Kubo formula, hence  
    $ \omega_E = 0$ in Eqs. (\ref{condLw}), (\ref{derspec}). In the case of  \Fig{figure8} the non-linear dc-electric field  effects are  included 
  using the solution to the classical equations of motion  with both ac- and dc-electric fields  (\ref{soleqm}).  
   
\begin{figure} [hbt]
\begin{center}
\includegraphics[width=4.5in]{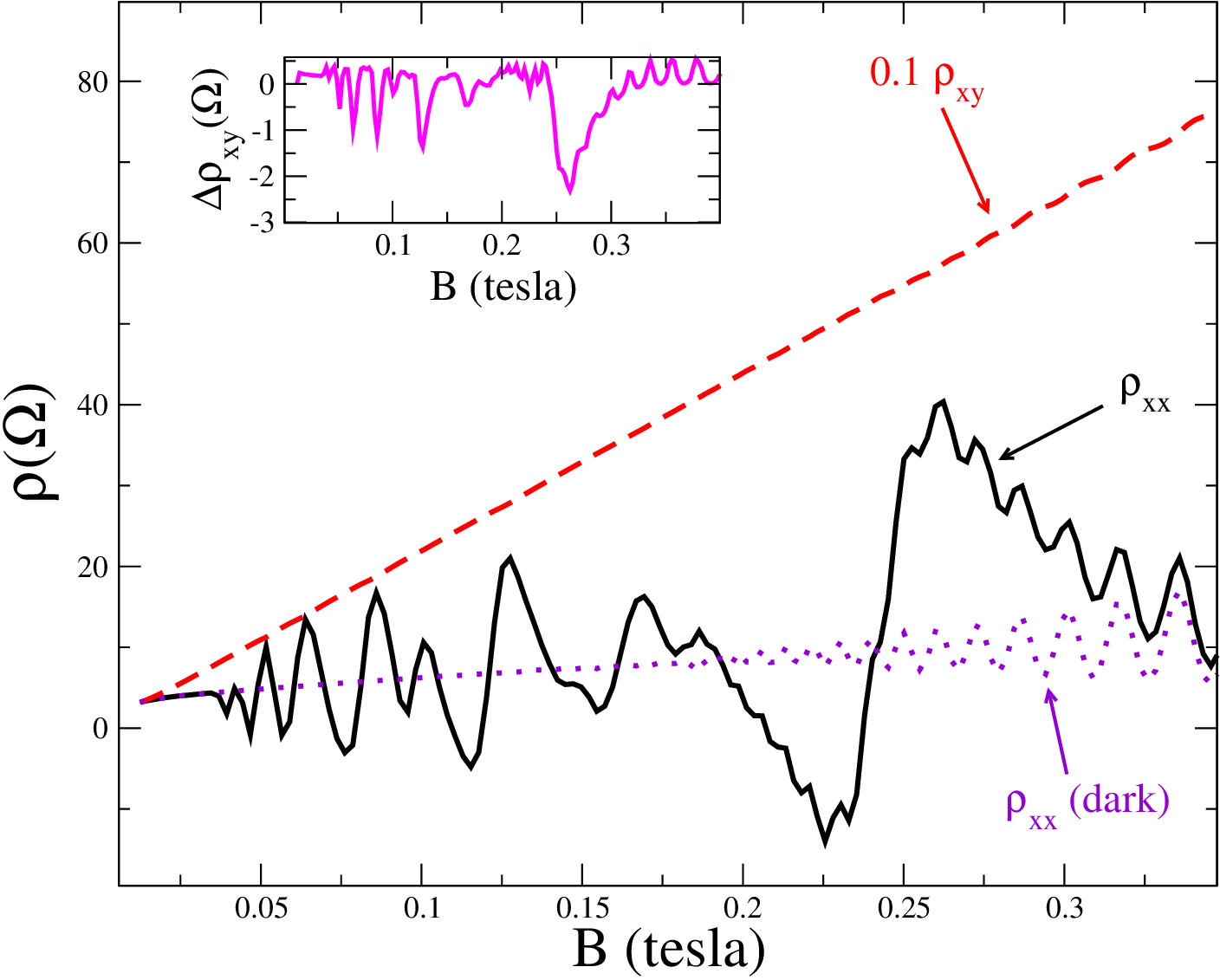}
\end{center}
\caption{ Magnetoresistivity $ \rho_{xx} $ versus $B$ without radiation and modulation (dotted line) and under microwave 
radiation plus periodic modulation (solid line). The dashed line corresponds to the total Hall resistance $ \rho_{xy} $, whereas the inset 
shows the results for the subtracted  Hall resistance $\Delta \rho_{xy} = \rho_{xy} - \rho_{xy}^{dark}$. 
 The microwave  polarization  is linear transverse (with respect to the current),  with  $f  = 100  \, Ghz$   and 
$\alpha = c \epsilon_0 \vert E_\omega \vert^2/(m^* \, \omega^3)$ = 0.4.  The other parameters are selected as follows:
$a = 25 \, nm$,  $V_0=0.05 \, meV $,
$m^* = 0.067 \, m_e$,  $ \mu \approx 2.5 \times 10^7 cm^2/V  s$,   
$\epsilon_F = 10 \, meV$, $T =1  \,  K$.   }
\label{figure1}
\end{figure}

   Plots of the  longitudinal and Hall resistivities as a function of the magnetic field intensity are displayed in \Fig{figure1}. The total  longitudinal resistance shows a strong oscillatory behavior 
   with distinctive NRS, this  behavior  is contrasted with the dark contributions that presents only the expected Shubnikov-de-Hass oscillations. The total Hall resistance presents a
   monotonous behavior, yet  perceptible microwave induced oscillations in the Hall 
   effect can be observed if one considers  $\Delta \rho_{xy}  =   \rho_{xy}  -  \rho_{xy}^{dark}$, see the inset.

\begin{figure} [hbt]
\begin{center}
\includegraphics[width=4.5in]{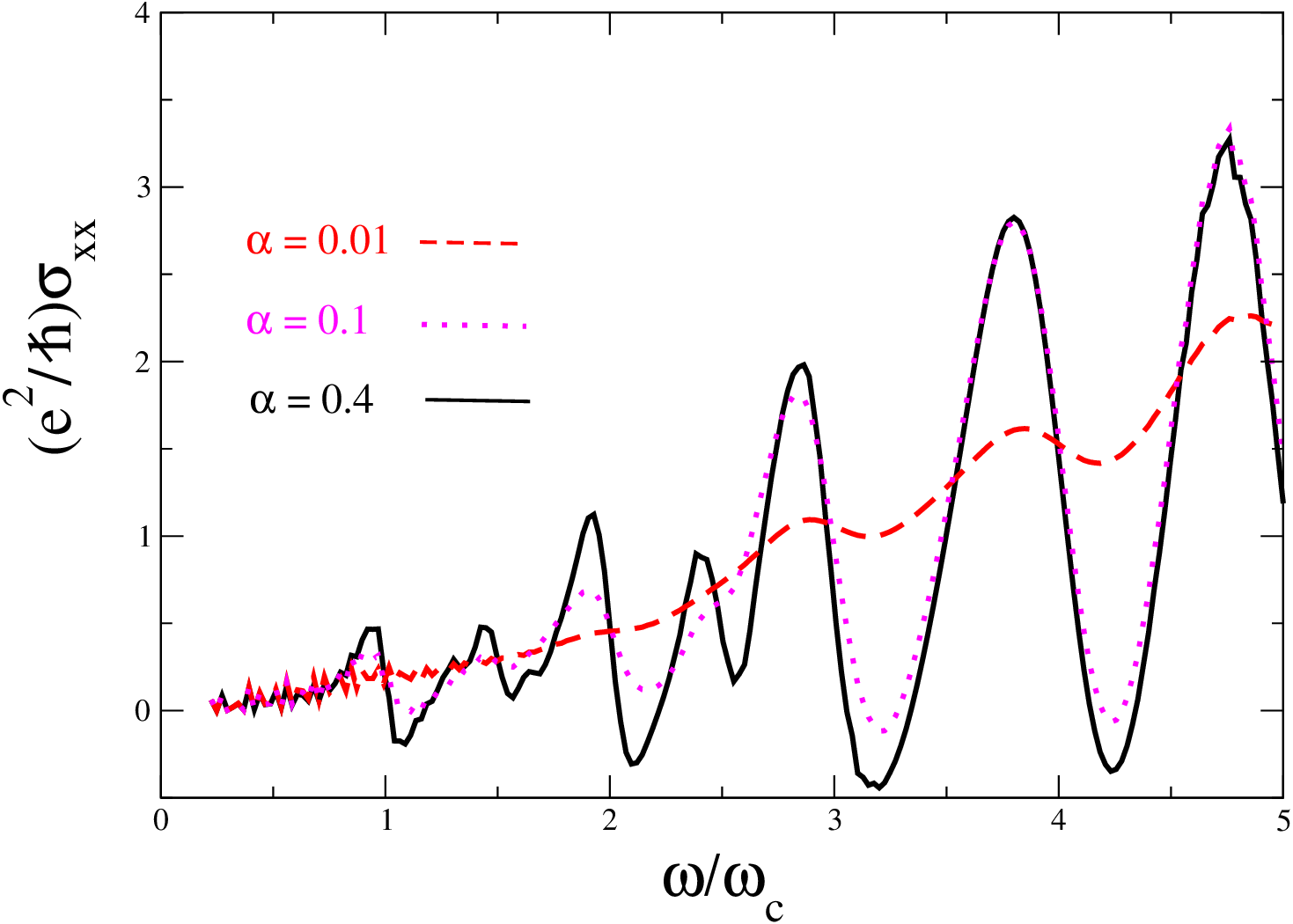}
\end{center}
\caption{ Longitudinal conductivity    versus $\epsilon =\omega/\omega_c$ for three  values of the microwave power intensity:   $\alpha = 0.01$  dashed line,  $\alpha = 0.1$ dotted line, and  $\alpha = 0.4$  continuos line. 
The other parameters 
  have the same values as in   \Fig{figure1}.  
}
\label{figure2}
\end{figure}

   \Fig{figure2} displays plots of the total longitudinal conductivity   as a function of   $\epsilon = \omega/ \omega_c  $. 
   $\sigma_{xx}$  shows  a strong oscillatory behavior,    
   with distinctive negative conductance   states. The periodicity as well as the  number of NCS  depend on the 
intensity of the   microwave radiation. For weak microwave intensity  ($\alpha =0.01 $), $\sigma_{xx}$ is positive with 
a moderate oscillatory behavior.   As the microwave intensity increases ($\alpha =0.1 $), strong oscillations in 
 $\sigma_{xx}$  are observed with   minima  centered
at    $\epsilon \sim 1.2$ , $\epsilon \sim 2.2$,  $\epsilon \sim 3.2$, and $\epsilon \sim 4.2$; only the two last minima correspond to NCS.  A further increase in the microwave intensity ($\alpha = 0.4$) yields  several NCS.  In the region $\epsilon \le  3$ the oscillation period is reduced to 
$\epsilon= \frac{1}{2}  $.
In general  it is observed  that $\sigma_{xx}$ vanishes at $\omega /\omega_c =j$ for $j$ integer.    The oscillations follow a pattern with minima
centered  at $\omega/\omega_c =j + \frac{1}{2} (l-1) +  \delta $,  and maxima centered at  $\omega/\omega_c =j + \frac{1}{2} (l-1) -  \delta $, where $j=1,2,3.......$,   $\delta \approx 1/5 $,  and $l$ is the dominant multipole that contributes to the conductivity in \Eq{condLw} . For moderate microwave  power the  $l=1$ ``one photon'' stimulated  processes dominate, corresponding to what is observed for  $\alpha =0.1 $. For $\alpha = 0.4$  the results can be interpreted as the results of  ``one and two  photon''  processes 
($l=1$ and $l =2$). To understand the origin of NCS  it is noticed from   \Eq{condLw}  that,  
for small microwave power, $\sigma_{xx}^\omega$ is dominated by   the $l=0$ Bessel term, that
  is always positive. Negative conductance states arise when the    $l=1$ and   $l=0$ terms  become 
comparable:  $ \vert  J_0\left( \vert \Delta  \vert \right) \vert^2  B^{(0)}  \sim \vert  J_1\left( \vert \Delta  \vert \right) \vert^2  B^{(1)}$. A simple analysis 
show that this condition is fulfilled for $ \vert \Delta  \vert \sim  0.1$. Using the expression  in \Eq{defdelta2},  the condition to produce 
NCS can be estimated  as $ \vert \mathbf{E}_\omega \vert  > E_{th}$ where $ E_{th} \approx 0.1\,  a \,  \Gamma_{rad} / \sqrt{8} e l_B $.   For the parameter used in   \Fig{figure2},    $ E_{th} \approx 10 \, V /cm $ or $\alpha_{th}  \approx 0.15$,  in good agreement with the results  displayed  by the plots.  

\begin{figure} [hbt]
\begin{center}
\includegraphics[width=4.5in]{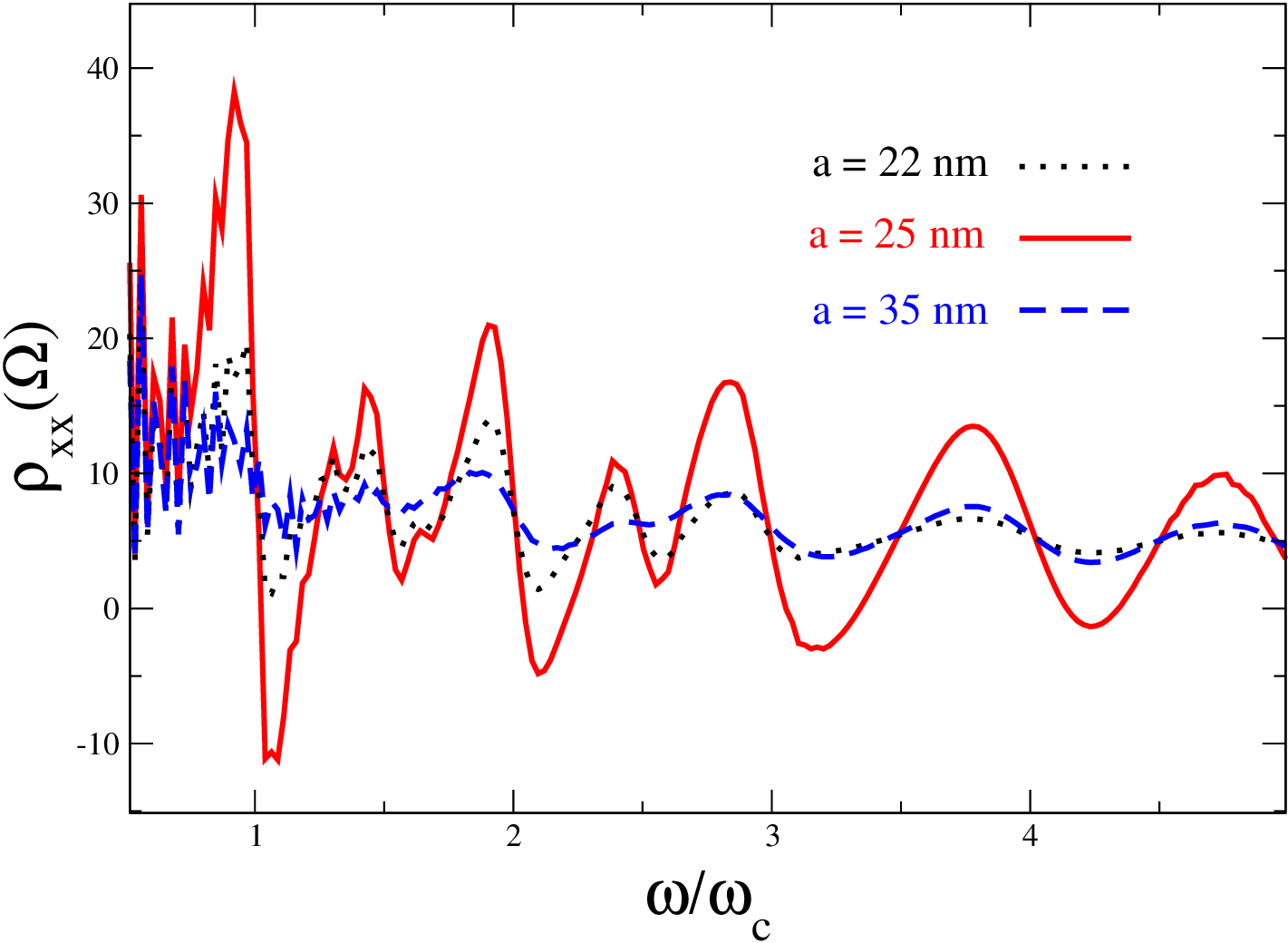}
\end{center}
\caption{  Longitudinal resistivity   versus $\epsilon =\omega/\omega_c$ 
for three  values of the lattice parameter:   $a=22 \, nm$ dotted line,  $a=25 \, nm$  continuos line,  and  $a=35 \, nm$    dashed  line. The microwave intensity is  $\alpha = 0.4$  and the other parameters 
  have the same values as in   \Fig{figure1}.  
  }
\label{figure3}
\end{figure}

Next we consider the dependence of $\rho_{xx}$ on the lattice parameter $a$.   Plots of  $\rho_{xx}$
versus  $\epsilon = \omega/ \omega_c  $ for various selections  of $a$  are presented  in   (\Fig{figure3}).
 NRS appear only in a narrow 
window of values of $a$ around   $a^*$, for which  the oscillations amplitude of $\rho_{xx}$  attains  its maximum. In the present case: 
$a^*  \approx 25 \, nm$. 
This behavior follows from the structure of the MM induced longitudinal conductance
$\sigma_{xx}^{MM}$ given in (\ref{condLw}).  Taking into account the form  of
$D_{\mu\nu}(\tilde{q}_{mn})$ in \Eq{laguerres}, the leading  dependence of $\sigma_{xx}^{MM}$  on the lattice parameter is given  approximately  by  $\sigma_{xx}^{MM} \sim \frac{1}{a^4} \exp \left( - \vert \tilde{q} \vert ^2 /2\right)$. According to  \Eq{psemom} $\tilde{q}  \propto 1/a$, consequently  the MM contribution is significant  only when $a$ is near to $a^*$, that
 is determined as:  $a^* \sim \pi l_B / \sqrt{2} $.  This estimation is in very good  agreement  with the numerical results presented in the plots.

Negative magnetoresistance requires ultra-clean  samples,  the phenomenon  appears when the 
     electron mobility exceeds  a threshold  $ \mu_{th}$.  \Fig{figure4}  displays $\rho_{xx} \, \, v.s. \,\, \omega / \omega_c $ plots for three  selected values of $\mu$. For $\mu \approx 0.5  \times 10^7 cm^2/V  s$  an   almost linear behavior $\rho_{xx} $ is clearly depicted (except in the Shubnikov-deHass region).  As the electron mobility  increases to $\mu \approx 1.5  \times 10^7 cm^2/V  s$, the resistance  oscillations are clearly observed; however,  several NRS  only  appear when   the mobility is increased to $\mu \approx 2.5  \times 10^7 cm^2/V  s$.  Eqs. (\ref{condLw},\ref{derspec}) contain the main ingredients that explain the huge increase observed in  the longitudinal  conductance (and resistance),   when the periodically modulated system is irradiated by microwaves and its critical dependence on the electron mobility.  In the standard expression for the Kubo formula there are no Floquet  replica contribution, hence $\omega$   can be set to zero in (\ref{derspec}), if that is   the case $B^{(l)}$ becomes proportional  to the energy derivative  of the Fermi distribution, that  in the $T \to 0$ limit  becomes of the form $\delta ({\cal E}  - {\cal E} _F)$, and the conductivity is positive definite  depending  only on  those states lying  at the Fermi level.  On the other hand,  as a result of the periodic structure induced by the  microwave radiation,  $B^{(l)}$  contains a second  contribution proportional to  the derivative of the density of states:   $   \frac{d}{d{\cal E} }  Im \, G_\nu ( {\cal E}  + l \omega )$.   Due to the oscillatory structure of the density of states, this extra contribution takes both  positive and negative values. According to \Eq{denst} this second term (as compared to the first one) is proportional to  the electron mobility, hence  for sufficiently  high mobility the new contribution dominates leading to  negative conductance states.

   The model is  used  to test  chirality effects induced by the magnetic field.    \Fig{figure5}   shows 
   $\sigma_{xx} \, \, v.s. \,\, \omega / \omega_c $ plots  for various   
   $\mathbf{E}_\omega$ field polarization's with respect to the current.   It is observed that  the   amplitude of  the $\sigma_{xx}$  oscillations    are    bigger for linear  transverse  polarization as compared to the  longitudinal polarization case.
  Similarly,  the oscillation amplitudes are enhanced  for negative circular polarization as compared to the
  positive circular polarization results; in particular NCS are  observed only for negative circular polarization.
   These results are understood  recalling  that   for negative circular polarization and $ \omega \approx \omega_c$ the  electric field rotates in phase with respect to the  electron cyclotron rotation. 

 \Fig{figure6} illustrates the fact that the strong   oscillations in $\sigma_{xx}$    originates from  the  combined
microwave-modulation  effects.
  The dark contribution,  \Eq{sigxxd}, shows the expected $\sigma_{xx}^D \propto B$
  linear behavior.  The contribution  arising solely from the periodic modulation $\sigma_{xx}^M$ can be isolated 
 from \Eq{condLw}  by switching-off the microwave field, the dashed line shows a smooth behavior. The  combined microwave-modulation contribution  is obtained 
 from $\sigma_{xx}^{MM} - \sigma_{xx}^M$; the dotted line clearly shows that this
 contribution includes the main oscillatory behavior of the full $\sigma_{xx}$, furthermore it is the only contribution that  becomes  negative. Hence, it is concluded that both the periodic modulation and the microwave radiation are essential  in order 
 to observe the strong  $\sigma_{xx}$ oscillations and the NCS.

     Next we explore the behavior of the longitudinal conductivity as a function of the microwave radiation intensity.
    As the intensity of the electric   microwave field ($E_\omega$)  is increased     higher   multipole  ($l$) terms needs  to be evaluated;  in the explored regime   convergent results are obtained  including terms up to the $l=5$ multipole.  In \Fig{figure7}  results are presented   for $\sigma_{xx}$  $vs$  $E_\omega$;  the selected values  of $\epsilon= \omega /\omega_c$  correspond to  minima or maxima of $\sigma_{xx}$ in figure
   \ref{figure2}.  In general it is observed that  for values corresponding to maxima $i.e.$  $\epsilon = j - \delta \,; \hskip0.5cm j= 1,2,3...$,  $\sigma_{xx}$ remains positive  for all    microwave field  intensities. 
   On the other hand, the minima corresponding to $\epsilon = j + \delta \,; \hskip0.5cm j= 1,2,3...$  
   are related to NCS around a region  $\vert E_\omega \vert  \in \left[10,25\right] \, V/cm $. Increasing the microwave 
   intensity leads to the disappearance  of the NCS, except for the first minima ($\epsilon \sim 1.1$).

  The non-linear regime with respect to the dc-external  field can also be explored within the present formalism. The effect is included 
  using the solution to the classical equations of motion  with both ac- and dc-electric fields, \Eq{soleqm}.  
   A possible connection between the observed  $ZRS$ in  $GaAs/Al_xGa_{1-x} As$ heterostructures  \cite{zudov1,mani1,zudov2,mani2} and the  the predicted  NRS\cite{ry1,ry2,durst,andre,shi,lei,vavilov,tor1}   was put forward by Andreev  $etal.$  \cite{andre}, noting that a general analysis of Maxwell equations shows that $NRS$  induces an instability  that  drives  the system  into a  ZRS. This mechanism requires  the longitudinal current $ j _{xx}$  as a function of $E_{dc}$ to have a single minimum, the system  instability will evolve  to the value $E_{dc}$ in which $ j _{xx}$ cancel.  Returning to the 
   irradiated superlattice  case, in \Fig{figure8} 
   it is observed that in general the  $ j _{xx}$ $vs.$  $E_{dc}$ plot has an oscillatory behavior, with more than one minima.  Hence   the conditions of the Andreeev  mechanism do not apply. Consequently, negative 
   conductance states may be probably observed in 2-dimensional   superlattices, when  exposed to both  magnetic and microwave fields.

 \section{Conclusions.}\label{conclu}

We have considered  a model to describe the microwave  photoconductivity of a  2DES in the presence of a  magnetic field, and   a 2D periodic modulation.  We presented a thoroughly discussion of the  method  that  takes into account  the Landau 
and microwave contributions in a non-perturbative exact way, the periodic potential effects are treated perturbatively. The formalism  exploits  the   symmetries  of the problem:  the exact  solution of the  Landau-microwave dynamics (\ref{tran1}) is obtained in terms of the electric-magnetic generators (\ref{trancan})  as well as the solutions  to the  classical  equations of motion (\ref{lagrang}).  The  spectrum  and Floquet modes are explicitly worked out (\ref{enerflo}).   In our model,  the Landau-Floquet  states act coherently with respect to the  oscillating field of  the superlattice potential, that   in turn induces  transitions between these levels.  Based on this formalism, a  Kubo-like   formula is provided (\ref{condLw})  that consistently  takes into account the oscillatory Floquet structure of the problem.

\begin{figure} [hbt]
\begin{center}
\includegraphics[width=4.5in]{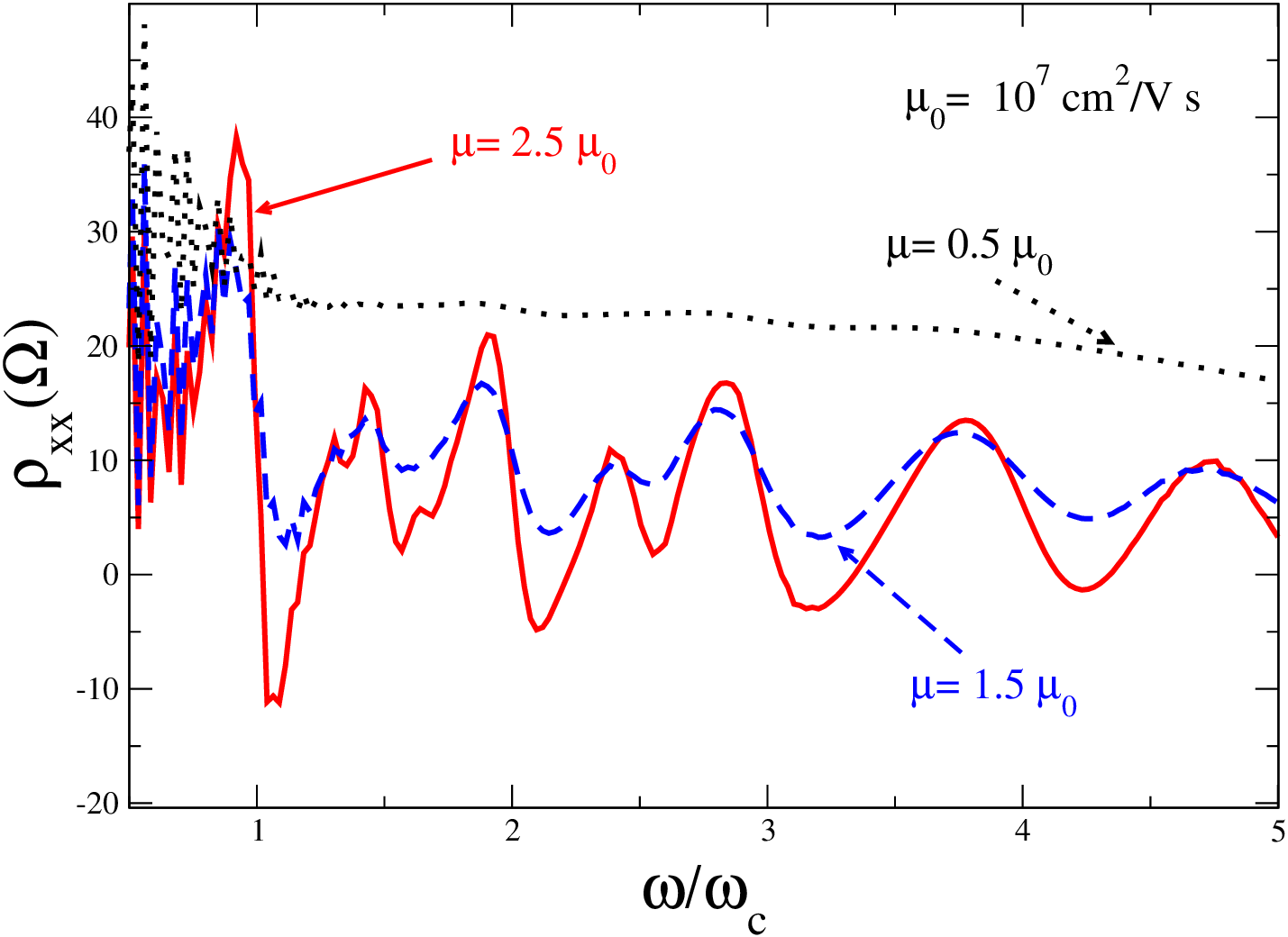}
\end{center}
\caption{ Longitudinal resistivity  as a function $\epsilon =\omega/\omega_c$ for three  values
of the electron mobility: $\mu = 0.5 \times 10^7 \, cm^2/Vs  $   dotted line, $\mu = 1.5 \times 10^7 \, cm^2/Vs  $ dashed line, and 
 $\mu = 2.5 \times 10^7 \, cm^2/Vs  $  continuos line. The microwave power is given by  $\alpha =  0.4$,  the other parameters 
  have the same values as in   \Fig{figure1}. 
 }
\label{figure4}
\end{figure}

\begin{figure} [hbt]
\begin{center}
\includegraphics[width=4.5in]{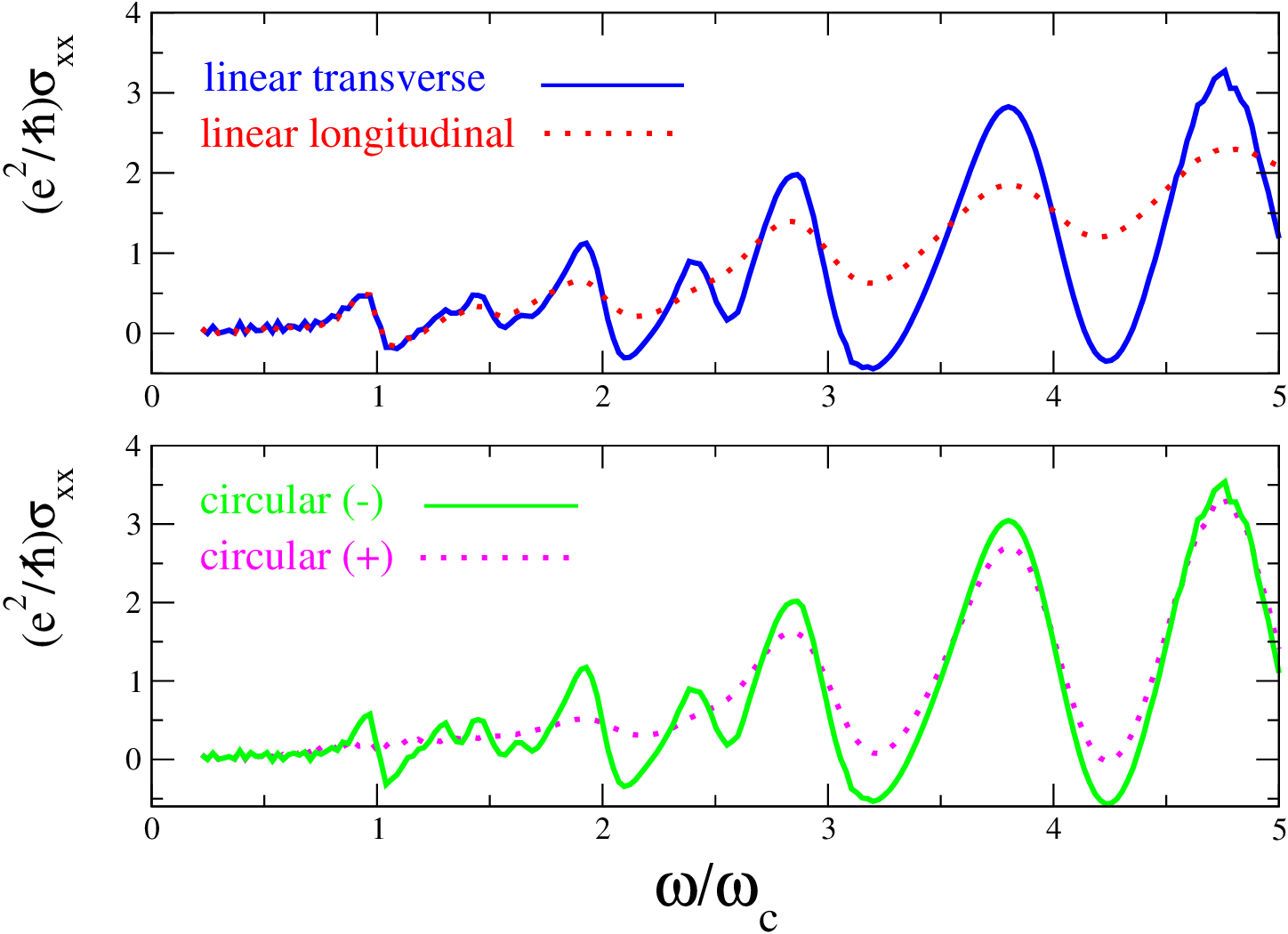}
\end{center}
\caption{ Longitudinal conductance $vs.$  $\epsilon =\omega/\omega_c$  for   various   microwave $E_\omega-$field polarization's with respect to the current.  In figure $(a)$ the continuos and dotted  lines correspond to  linear   transverse and longitudinal  polarization's   respectively. 
Figure $(b)$  shows results for circular polarization's:  left-hand (continuos line) and  right-hand  (dotted line). $\alpha =  4$ and  the values of the other  parameters are the same as in figure \Fig{figure1}
 }
\label{figure5}
\end{figure}

\begin{figure} [hbt]
\begin{center}
\includegraphics[width=4.5in]{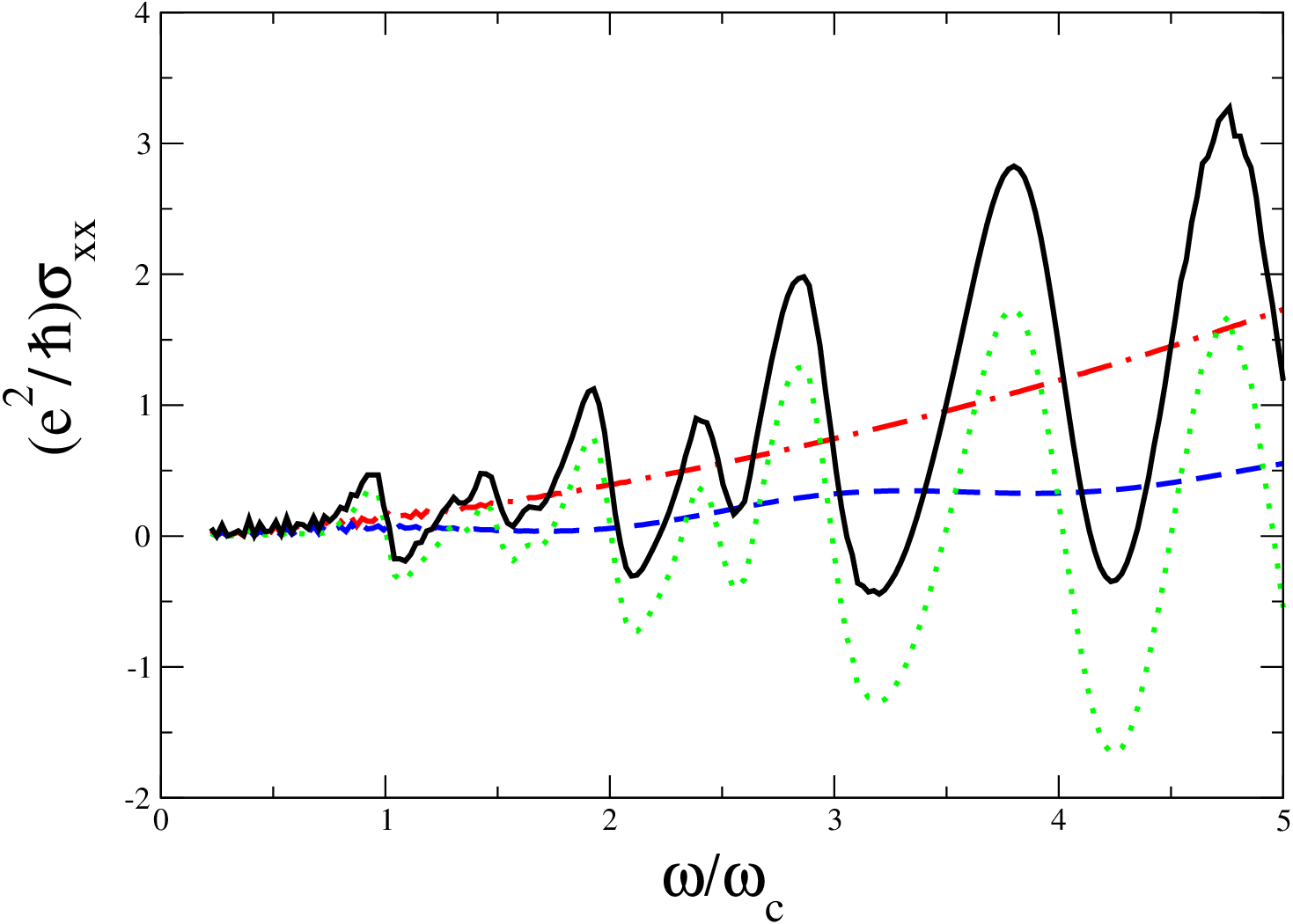}
\end{center}
\caption{ Contributions to the total $\sigma_{xx}$  (continuos line)  $vs.$  $\epsilon =\omega/\omega_c$: dark contribution $\sigma_{xx}^D $ (dashed-dotted line) from \Eq{sigxxd},  periodic modulation 
contribution $\sigma_{xx}^M$ (dashed line) obtained from \Eq{condLw} when the microwave field is switched-off, microwave-modulation contribution obtained as 
$\sigma_{xx}^{MM} -  \sigma_{xx}^{M}$ .  The values of the other  parameters are the same as in figure \Fig{figure1}
 }
\label{figure6}
\end{figure}

\begin{figure} [hbt]
\begin{center}
\includegraphics[width=4.5in]{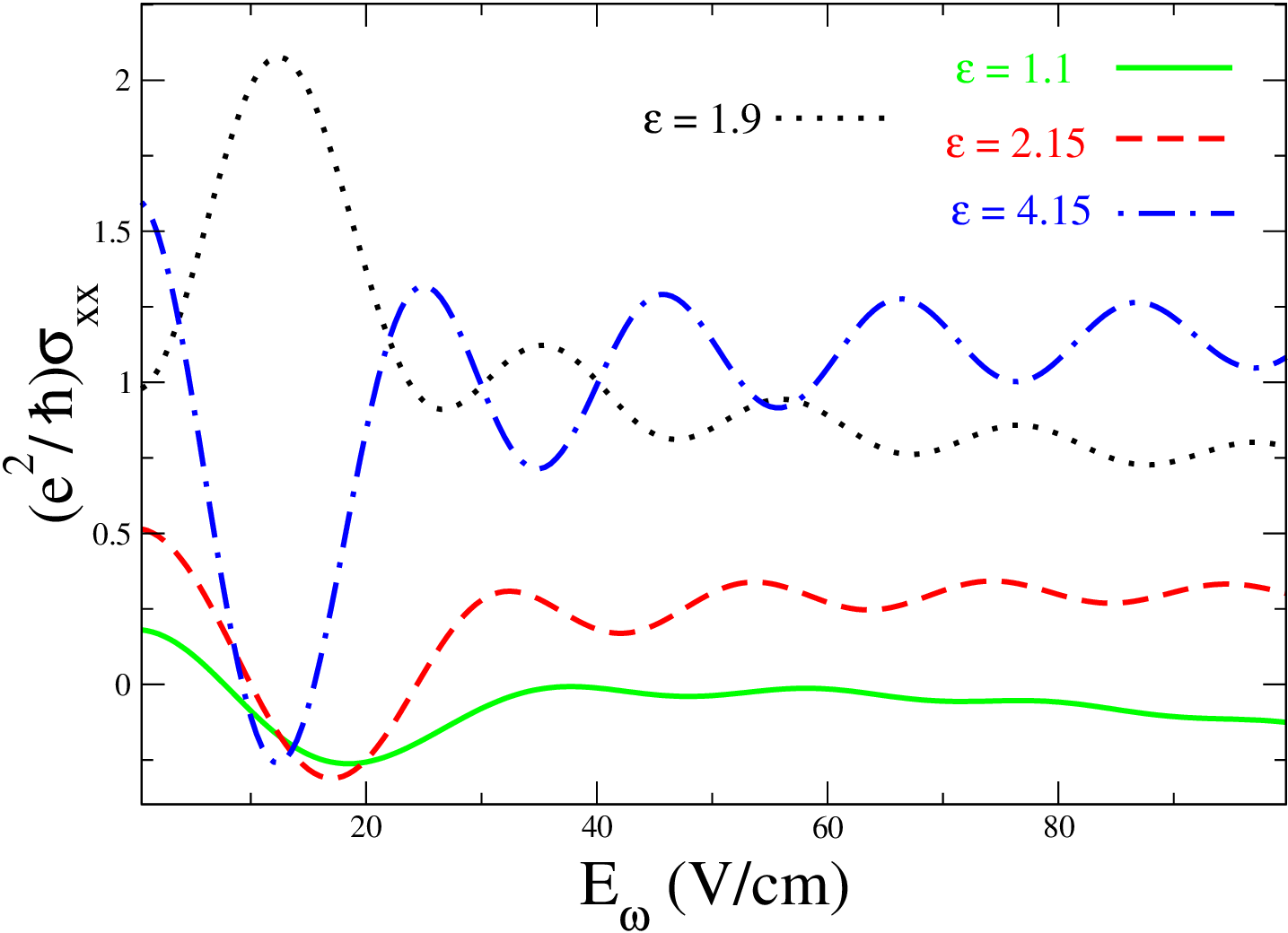}
\end{center}
\caption{ Longitudinal conductivity as a function of  the microwave ac-electric field      for various  values  of $\epsilon = \omega/\omega_c$: 
$\epsilon=1.1 $ continuos line,  $\epsilon=1.9 $ dotted line, $\epsilon=2.15 $ dashed line, and $\epsilon=4.15 $  dashed-dotted line. 
 The other parameters 
  have the same values as in   \Fig{figure1} }
\label{figure7}
\end{figure}

\begin{figure} [hbt]
\begin{center}
\includegraphics[width=4.5in]{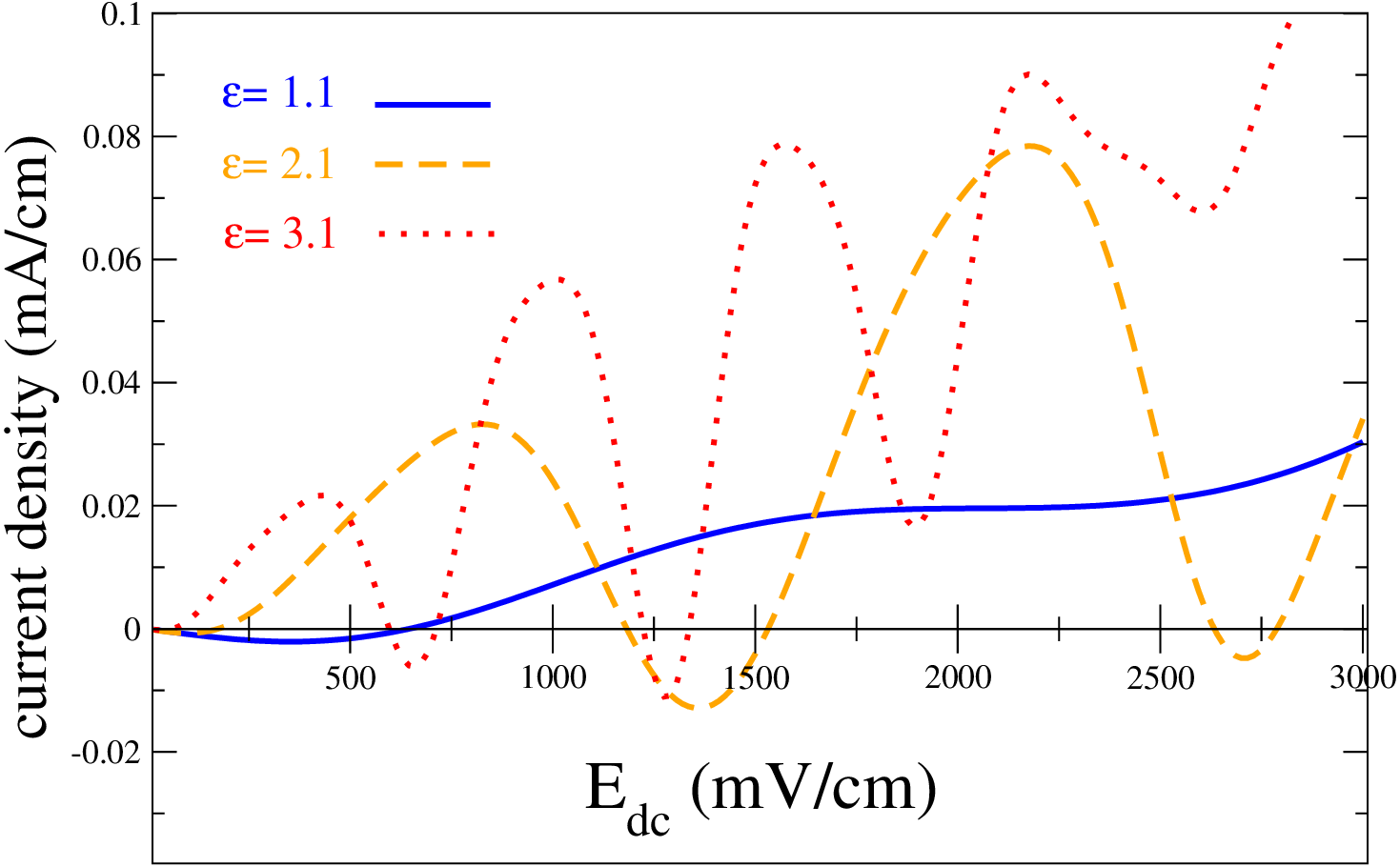}
\end{center}
\caption{  Current-voltage characteristics for the irradiated sample 
for various  values 
of $\epsilon = \omega/\omega_c$: 
$\epsilon=1.1 $ continuos line,  $\epsilon=2.1 $ dashed line, and $\epsilon=3.15 $  dotted line. 
The microwave power is given by  $\alpha =  0.4$,  the other parameters 
  have the same values as in   \Fig{figure1}
 }
\label{figure8}
\end{figure}

 It is found  that  both  $\sigma_{xx}$ and $\rho_{xx}$  exhibit strong oscillations governed  by   $\epsilon = \omega /  \omega_c$. 
The oscillations follow a pattern with minima
centered  at $\omega/\omega_c =j + \frac{1}{2} (l-1) +  \delta $,  and maxima centered at  $\omega/\omega_c =j + \frac{1}{2} (l-1) -  \delta $, where $j=1,2,3.......$,   $\delta \approx 1/5$  and $l$ is the dominant multipole  contribution. 
 NRS develop for sufficiently  strong  microwave power (\Fig{figure2}),  in a narrow window of values of the lattice parameter $(a \sim l_B)$   (\Fig{figure3}),   and for  high  electron mobility samples
 (\Fig{figure4}). The explanation for the NRS can be traced down to 
 Eqs.    (\ref{condLw}) and  (\ref{derspec});  the  longitudinal photoconductivity contains a  new contribution proportional to the derivative of the density of states:
   $   \frac{d}{d{\cal E} }  Im \, G_\nu ( {\cal E}  + l \omega )$.   Due to the oscillatory structure of the density of states this extra contribution takes both  positive and negative values. This  term is proportional to the electron mobility, hence  for sufficiently  high mobility the new contribution dominates leading to  negative conductivity  states.  Unlike the  semiclassical origin of magnetoresistance oscillations observed in  an antidot array for 
   commensurate values of the ratio $R_c/a$ \cite{vassi}, these  conductance oscillations have a quantum origin and would  only appear in a narrow window of values of $a$, around $ a \sim l_B$.

In conclusion, it is  proposed that 
 the   combined effect of: periodic modulation,  perpendicular magnetic field,  plus microwave  irradiation   of  2DES give rise to interesting oscillatory conductance phenomena, with the possible   development of NCS and NRS.  One should stress  that according to our results,  the production of NRS requires ultra-clean samples with electron mobilities of order  $\mu \approx 2.5  \times 10^7 cm^2/V  s$ (see \Fig{figure4}). The electron mobilities 
in the  fabricated  arrays of periodic scatterers so far  \cite{klitbut,vassi} are $\mu \approx  2.5  \times 10^6 cm^2/V  s$,  consequently 
an increase on the electron mobilities  of these kind of experimental setups  by an order of magnitude would be required in order to observe the phenomena described in this work.


\subsection{Appendix  A: Dark and Hall  conductivities.}\label{apenA}

   In section (\ref{kubo}) it was explained in detail the method  to obtain the final expression for the 
 modulation-microwave induced conductance  \Eq{condLw}. 
   Working along a similar procedure the expression for the  remaining conductivities are worked  from  
    equations (\ref{condd})  and (\ref{condw}).  First we quote the 
  longitudinal  dark conductance 
      \begin{equation}\label{sigxxd}
 \mathbf{\sigma}^{D}_{xx}   =  \frac{ e^2 \omega_c^2}{\pi \hbar}   \sum_\mu \, \mu \int d{\cal E}  \,  Im \, G_\mu\left( {\cal E}   \right)
 \,   \frac{d f }{d {\cal E}  } \,  Im \, G_\mu \left( {\cal E}  + \omega_c \right) \, , 
 \end{equation}
 whereas the  Hall conductance  reads 
\hskip-2.0cm    \begin{equation}\label{sigxyd}
 \mathbf{\sigma}^{D}_{xy}   =  \frac{ e^2 \omega_c^2 }{\pi \hbar  } \sum_\mu  \, \mu \int d{\cal E}  \, Im \, G_\mu\left( {\cal E}   \right)
 \left[ f\left( {\cal E} _\mu - \omega_c \right) -   f\left( {\cal E}  \right)\right]  \, P\, \frac{1}{\left({\cal E}  - {\cal E} _\mu + \omega_c \right)^2}
 \, ,   
 \end{equation}
 where $P$ indicates the principal-value integral. 
   The  final result for the MM  assisted 
 longitudinal conductivity was quoted in \Eq{condLw}.  The MM induced
  Hal conductivity is calculated to give 
 \begin{equation}\label{sigxyw}
\hskip-2.5cm   \mathbf{\sigma}_{xy} ^{(MM)}   =   \frac{ e^2 l_B^4 }{\pi \hbar }  
\int d{\cal E}     \sum_{\mu \nu}  \sum_l \sum_{mn}   Im \, G_\mu \left({\cal E}   \right)  
 \left[ f\left( {\cal E} _\nu  \right) -   f\left( {\cal E}  \right)\right] 
 T_{mn}^l \, \, \bigg{\vert}  J_l\left( \vert \Delta_{mn} \vert \right)  V_{mn}  D_{\mu\nu} (\tilde{q}_{mn}) \bigg{\vert}^2     ,
 \end{equation}
 were the function $T_{mn}^l$ is defined as
\begin{equation} \label{funaux2} 
T_{mn}^l = \omega_c^3 \,    \, \frac{  \left(q_m^{(x)})^2 \, + \,   (q_n^{(y)} \right)^2 }
   { \left(  {\cal E}  + \omega l  - {\cal E} _\nu    \right) \,\, {\vert \left({\cal E}  + \omega l -  {\cal E} _\nu    \right)^2 - \omega_c^2 \vert^2}} \, .   
 \end{equation}

\subsection{Appendix  B: Microwave-driven distribution function.}\label{apenB}

Within  the time relaxation approximation the Boltzmann equation  can be written as 
   \begin{equation}\label{boltz}
\frac{\partial f }{\partial t } + \frac{\partial f }{\partial  \mathbf{p}  } \cdot \left( e  \mathbf{E}  + e
 \mathbf{v}    \times  \mathbf{B}    \right) = - \frac{f - f_F }{\tau_{tr}} 
- \frac{f - f_F }{\tau_{in}}  ,
\nonumber 
 \end{equation}
 where  $f_F$  is  the Fermi-Dirac distribution and  we distinguish between the elastic  rate $\tau_{tr}^{-1}$ and inelastic or energy relaxation rate $\tau_{in}^{-1}$. 
 As it already mentioned, we assume the validity of the following  conditions:  $\tau_\omega \ll \tau_{tr} \ll  \tau_{in} $, and certainly the inelastic processes can be safely ignored. Furthermore,  due  to the $ac$-electric field (\ref{vecpot}),   the L.H. S. of the previous equation is estimated to be of order $f / \tau_\omega$; hence, in a first approximation  the elastic scattering contribution can also be neglected. The resulting Vlasov equation has the exact solution $f(  \mathbf{p}  , t) = f_F \left(  \mathbf{p}   - m^*  \mathbf{v}  (t)\right) $, where the velocity 
 $  \mathbf{v}  (t) \equiv \left(\dot \eta_1, \dot \zeta_1 \right) $ solves exactly the same classical equations of motion that follow from (\ref{lagrang}), and the initial condition is  selected 
 as,  $f  \to f_F$  as the external electric field is switched-off.  In particular it is  verified that 
 $ m^* \vert  \mathbf{v}   (t) \vert^2/2 =  {\cal E}_{rad} $  coincides with  the Floquet energy shift  produced by the microwave radiation (\ref{enerflo}). 
 The steady-state distribution,  evaluated at  the Landau energy ${\cal E} ={\cal E} _\mu^{(0)}$,  is obtained by averaging $ f_F \left(  \mathbf{p}   - m^*  \mathbf{v}  (t)\right) $ over the oscillatory period
   \begin{equation}\label{boltz2}
\langle f_F \rangle = \frac{1}{\tau_\omega} \int_0^{\tau_\omega} f_F \left( {\cal E} _\mu^{(0)}  +  {\cal E}_{rad}+
2 \, \cos \omega_c t   \,  \sqrt{ {\cal E} _\mu^{(0)} \,   {\cal E}_{rad}} \,  \right)  \, dt 
\nonumber . 
 \end{equation}
 In general  it is verified that  $  {\cal E}_{rad}  \ll {\cal E} _\mu^{(0)}  $, thus  expanding to first order 
one finds   $\langle f_F \rangle   \approx f_F \left( {\cal E} _\mu^{(0)}  +  {\cal E}_{rad} \right) =   f_F( {\cal E} _\mu   ) $. Hence, it is verified that a  rapid relaxation of the Fermi distribution to the quasi-energy states is a  reasonable assumption. The arguments presented in this appendix have  been introduced  by  Mikhailov \cite{mikha} in order to explore the possibility that the microwave radiation leads to a population inversion;  however,  it is concluded  that it  would require a rather high   microwave intensity 
 $  {\cal E}_{rad} > {\cal E} _F$.

\ack 
 We acknowledge the partial financial support endowed by
CONACyT through grants No.   \texttt{42026-F} and   \texttt{G32736-E}, and UNAM project No.  \texttt{IN113305}.

\newpage

\section*{References} 

\begin{thebibliography}{10}
  
\bibitem{zudov1}
M. A. Zudov, R. R. Du, J. A. Simmons, J. L. Reno, 
{\em  Phys. Rev. B \/}
{\bf  64}  
(2001)
201311(R).


\bibitem{zudov2}
M. A. Zudov, R. R. Du, L. N. Pfeiffer, K. W. West,
{\em  Phys. Rev. Lett \/}
{\bf  90}  
(2003)
046807.

\bibitem{mani1}
R. G. Mani, J. H. Smet, K. von klitzing, V. Narayanamurti, W.b. Johnson, Umansky,
{\em  Nature \/}
{\bf  420}  
(2002)
646;
{\em  Phys. Rev. Lett \/}
{\bf  92}  
(2004)
146801.


\bibitem{mani2}
R. G. Mani, 
{\em  Physica E\/}
{\bf  22}  
(2004)
1.


\bibitem{doro}
S. I. Dorozhkin,
{\em  JETP. Letters. \/}
{\bf  77}  
(2003)
577.


\bibitem{willett}
R. L.  Willett, L.  N. Pfeiffer, K. W. West,
{\em  Phys. Rev. Lett \/}
{\bf  93}  
(2004)
026804.

 \bibitem{zudov3}
M. A. Zudov, 
{\em  Phys. Rev. B  \/}
{\bf  69}  
(2004)
041304(R).

 \bibitem{kova}
A. E. Kovalev, S. A. Zvyagin, C. R.  Bowers, J. L. Reno, J. A. Simmons,
{\em   Solid State Commun. \/}
{\bf  1130}  
(2004)
379.

\bibitem{stud}
S. A. Studenikin, M. Potemski, A. Sachrajda, M. Hilke, L. N. Pfeiffer, K. W. West,
{\em cond-mat/0404411 \/} (2004). 

 \bibitem{ry1}
V. I. Ryzhii,
{\em  Sov. Phys. Solid State \/}
{\bf  11}  
(1970)
2078.

\bibitem{ry2}
V. I. Ryzhii, R. Suris,
{\em  J Phys. Cond. Matt. \/}
{\bf  15}  
(2003)
6855.

\bibitem{durst}
A. C. Durst, S. Sachdev, N. Read, S. M. Girvin, 
{\em  Phys. Rev. Lett. \/}
{\bf  91}  
(2003)
086803.

\bibitem{andre}
A. V. Andreev, I. L. Aleiner, A. J. Millis, 
{\em  Phys. Rev. Lett. \/}
{\bf  91}  
(2003)
056803. 


\bibitem{shi}
J. Shi, X. C. Xie,  
{\em  Phys. Rev. Lett. \/}
{\bf  91}  
(2003)
086801.

\bibitem{lei}
X. L. Lei, S. Y. Liu,
{\em  Phys. Rev. Lett. \/}
{\bf  91}  
(2003)
226805.

 \bibitem{vavilov}
M. G. Vavilov, I. L. Aleiner,
{\em  Phys. Rev. B  \/}
{\bf  69}  
(2004)
035303.

 \bibitem{tor1}
M. Torres, A. Kunold,
{\em  Phys. Rev. B  \/}
{\bf  71}  
(2005)
115313-1.

 \bibitem{tor2}
M. Torres, A. Kunold,
{\em  Phys. Stat. Sol. b  \/}
{\bf  71}  
(2005)
1192.


\bibitem{dmi1}
I. A. Dmitriev, A. D. Mirlin, D. G. Polyakov,
{\em  Phys. Rev. Lett. \/}
{\bf  91 }  
(2003)
226802.

\bibitem{dmi2}
I. A. Dmitriev, M. G. Vavilov, I. L. Aleiner,  A. D. Mirlin, D. G. Polyakov,
 {\em Phys. Rev. B \/}
  {\bf 71}
 (2005)
115316. 


\bibitem{kennett}
J. P. Robinson, M. P. Kennett, , N. R. Cooper, V. I. Fal'ko,
{\em  Phys. Rev. Lett. \/}
{\bf  93}  
(2004)
036804

\bibitem{weiss-vk}
D. Weiss,  K. v. Klitzing, K. Ploog, G. Weimann,  
{\em  Europhys. Lett.   \/}
{\bf  8 }  
(1989)
179;
R. W. Winkler, J. P. Kotthaus, K. Ploog,  
{\em  Phys. Rev. Lett.  \/}
{\bf  62 }  
(1989)
1177;
D. Weiss, M. L. Roukes, A. Menschig, P. Grambow,  K. v. Klitzing, G. Weimann,  
{\em  Phys. Rev. Lett. . \/}
{\bf  66 }  
(1991)
2790.


\bibitem{schuster}
 For a review see R. Schuster, K. Ensslin,  
{\em  Adv. Solid State Phys.  \/}
{\bf  34 }  
(1994)
195.


\bibitem{klitbut}
 C. Albrecht, J. H. Smet,   K. v. Klitzing, D. Weiss, V. Umansky, H. Schweizer ,  
{\em  Phys. Rev. Lett.  \/}
{\bf  86 }  
(2001)
147.

\bibitem{vassi}
 E. Vasiliadou, R Fleischmann, D. Weiss, D.  Heitmann,    K. v. Klitzing,  T. Geisel , R.  Bergmann,
  H. Schweizer, C. T. Foxon,   
{\em  Phys. Rev.   \/}
{\bf  B 52  }  
(1995)
R8658.

\bibitem{dietel}
J. Dietel, L. I. Glazman, F. W. J. Hekking, F. von Oppen,  
{\em  Phys. Rev. B \/}
{\bf  71}  
(2005)
045329.

\bibitem{gumbs}
G. Gumbs 
{\em  Phys. Rev. B \/}
{\bf  72}  
(2005)
125342.

\bibitem{husi}
K. Husimi, 
{\em  Prog. Theor. Phys.  \/}
{\bf  9}  
(1953)
381. 

\bibitem{Kunold1}
A. Kunold, M. Torres,
{ \em Phys. Rev. B \/}
 {\bf 61 }
(2000)
9879.

\bibitem{Torres1}
   M. Torres, A. Kunold,
{ \em Phys. Lett. A\/}
 {\bf 323 }
(2004)
2890.

\bibitem{Ashby1}
N. Ashby and S.C. Miller,
 {\em Phys. Rev. B \/}
  {\bf 139 }
 (1965)
A428. 

\bibitem{Kunold2}
  A. Kunold, M. Torres, 
{\em Annals of Physics \/}
  {\bf 315 }
 (2005)
532. 


\bibitem{ando1}
  T. Ando, Y. Uemura, 
 {\em J. Phys. Soc. Japan.  \/}
  {\bf 36 }
 (1974)
959.

\bibitem{gerh}
  R. R. Gerhardts, 
 {\em Z. Phys. B  \/}
  {\bf 21 }
 (1975)
275;
Ibid. 
{\bf 21 }
 (1975)
285. 


\bibitem{ando2}
  T. Ando, A. B. Fowler, F. Stern, 
 {\em Rev. Mod. Phys.  \/}
  {\bf 54 }
 (1982)
437.


\bibitem{quant}
  T. Dittrich, P. hanggi, G.L. Ingold, B. Kramer, G. Schon, W. Zwerger, {\it ``Quantum transport and dissipation''},  Wiley-VCH
 (1998).


\bibitem{mikha}
  S. A. Mikhailov,
 {\em Phys. Rev. B \/}
  {\bf 70 }
 (2004)
165311. 

 \end{thebibliography}
\end{document}